\documentclass{aa} 

\usepackage{pifont}
\usepackage{graphicx}
\usepackage{txfonts}
\usepackage{CJK}
\usepackage{enumitem}
\usepackage{amsmath}
\usepackage{amssymb}
\usepackage{natbib}
\bibpunct{(}{)}{;}{a}{}{,}
\usepackage{url}
\usepackage{marvosym}
\usepackage{hyperref}
\usepackage{xcolor}
\hypersetup{
    colorlinks=true,
    linkcolor=blue,
    filecolor=blue,      
    urlcolor=blue,
    citecolor=blue
}
\usepackage{multirow}

\newcommand{\EBV}{E(B\,{-}\,V)}
\newcommand{\EBPRP}{E(BP\,{-}\,RP)}
\newcommand{\Av}{A_{\rm V}}
\newcommand{\teff}{T_{\rm eff}}
\newcommand{\logg}{\log{g}}
\newcommand{\meta}{{\rm [M/H]}}
\newcommand{\gspspec}{GSP-Spec}
\newcommand{\dibdepth}{\mathcal{D}}
\newcommand{\diblambda}{\lambda_{\rm DIB}}
\newcommand{\dibwidth}{\sigma_{\rm DIB}}
\newcommand{\Vdib}{V_{\rm DIB}}
\newcommand{\Vstar}{V_{\rm star}}
\newcommand{\kms}{\rm km\,s^{-1}}
\newcommand{\NHI}{N_\ion{H}{i}}
\newcommand{\NHt}{N_{\rm H_2}}

\defcitealias{Saydjari2023}{AS23}
\defcitealias{Schultheis2023DR3}{GDR3}
\defcitealias{Schultheis2023FPR}{GFPR}

\makeatletter
\renewcommand*\aa@pageof{, page \thepage{} of \pageref*{LastPage}}
\makeatother

\begin{document}

\begin{CJK*}{UTF8}{gbsn}

\title{Diffuse Interstellar Bands in Gaia DR3 RVS spectra}
\subtitle{Machine-learning based new measurements}

\author{H. Zhao (赵赫)\inst{1} \thanks{Corresponding author: He Zhao}
        \and
        M. Schultheis\inst{2}
        \and
        C. Qu (屈彩霞)\inst{3,4}
        \and 
        T. Zwitter\inst{5}
        }

\institute{Purple Mountain Observatory and Key Laboratory of Radio Astronomy, Chinese Academy of Sciences, 10 Yuanhua Road, Nanjing 210033, People's Republic of China \\
          \email{hzhao@pmo.ac.cn, mathias.schultheis@oca.eu} 
          \and
          University C\^ote d'Azur, Observatory of the C\^ote d'Azur, CNRS, Lagrange
          Laboratory, Observatory Bd, CS 34229, \\
          06304 Nice cedex 4, France 
          \and
          CAS Key Laboratory of Optical Astronomy, National Astronomical Observatories, Chinese Academy of Sciences, Beijing 100101, China
          \and 
          School of Astronomy and Space Science, University of Chinese Academy of Sciences, Beijing 100049, China
          \and
          Faculty of Mathematics and Physics, University of Ljubljana, Jadranska 19, 1000 Ljubljana, Slovenia
          }

\date{Received 20 November 2023; accepted 15 December 2023}


 
\abstract
{Diffuse interstellar bands (DIBs) are weak and broad interstellar absorption features in astronomical spectra originating from 
unknown molecules. To measure DIBs in spectra of late-type stars more accurately and more efficiently, we developed a Random
Forest model to isolate the DIB features from the stellar components and applied this method to 780 thousand spectra collected by 
the Gaia Radial Velocity Spectrometer (RVS) that were published in the third data release (DR3). After subtracting the stellar
components, we modeled the DIB at 8621\,{\AA} ($\lambda$8621) with a Gaussian function and the DIB around 8648\,{\AA} ($\lambda$8648)
with a Lorentzian function. After quality control, we selected 7619 reliable measurements for DIB\,$\lambda$8621. The equivalent
width (EW) of DIB\,$\lambda$8621 presented a moderate linear correlation with dust reddening, which was consistent with our previous 
measurements in Gaia DR3 and the newly Focused Product Release. The rest-frame wavelength of DIB\,$\lambda$8621 was updated as 
$\lambda_0\,{=}\,8623.141\,{\pm}\,0.030$\,{\AA} in vacuum, corresponding to 8620.766\,{\AA} in air, which was determined by 77 DIB 
measurements toward the Galactic anti-center. The mean uncertainty of the fitted central wavelength of these 77 measurements is 
0.256\,{\AA}. With the peak finding method and a coarse analysis, DIB\,$\lambda$8621 was found to correlate better with the neutral 
hydrogen than the molecular hydrogen (represented by $^{12}$CO $J\,{=}\,(1{-}0)$ emission). We also obtained 179 reliable measurements 
of DIB\,$\lambda$8648 in the RVS spectra of individual stars for the first time, further confirming this very broad DIB feature. 
Its EW and central wavelength presented a linear relation with those of DIB\,$\lambda$8621. A rough estimation of $\lambda_0$ for 
DIB\,$\lambda$8648 was 8646.31\,{\AA} in vacuum, corresponding to 8643.93\,{\AA} in air, assuming that the carriers of $\lambda$8621 
and $\lambda$8648 are co-moving. Finally, we confirmed the impact of stellar residuals on the DIB measurements in Gaia DR3, which 
led to a distortion of the DIB profile and a shift of the center ($\lesssim$0.5\,{\AA}), but the EW was consistent with our new 
measurements. With our measurements and analyses, we propose that the machine-learning-based approach can be widely applied to 
measure DIBs in numerous spectra from spectroscopic surveys. 






}

\maketitle

\titlerunning{Gaia DIB}

\section{Introduction} \label{sect:intro}

Diffuse interstellar bands (DIBs) are a set of absorption features in the spectra of stars, galaxies, and quasars observed in the 
optical and near-infrared bands (about 0.4--2.4\,$\mu$m, see DIB surveys: \citealt{Fan2019}; \citealt{Hamano2022}; \citealt{Ebenbichler2022}).
High-quality astronomical observations, experimental measurements, and theoretical analysis support that the DIBs originate from 
complex carbon-bearing molecules \citep[e.g.][]{Campbell2015,Omont2019,MacIsaac2022}, so that DIBs can be served as chemical and
kinematic tracers of Galactic interstellar medium (ISM), despite the exact species of most DIB carriers are still unknown.

Because DIBs are weak features and could be blended with stellar lines, early studies preferred observing a handful of hot stars 
as background sources because of their clean spectra, which favored the survey of DIB signals \citep[e.g.][]{JD1994,Galazutdinov2000a,
Hobbs2008,Hobbs2009,Fan2019} and the exploration of elemental properties of DIB carriers \citep[e.g.][]{Friedman2011,Vos2011,Fan2017}. 
On the other hand, large spectroscopic surveys during the last decade, such as RAVE \citep{Steinmetz2006}, APOGEE \citep{Majewski2017}, 
GALAH \citep{Buder2021}, and Gaia Radial Velocity Spectrometer \citep[RVS;][]{Cropper2018,Sartoretti2018} have observed the spectra 
of hundreds of thousands to tens of millions of stars, which enables statistical studies of DIB properties. For example, \citet{Lan2015} 
and \citet{Baron2015a} mapped the DIB strength projected on the celestial sphere at high latitudes. \citet{Kos2014} and \citet{Zasowski2015c} 
built the three-dimensional (3D) distribution of DIB strength for DIBs at 8621\,{\AA} ($\lambda$8621) and 15273\,{\AA}, respectively. 
\citet{hz2021b} and \citet{Zasowski2015c} further investigated the kinematics of these two DIBs.

Because late-type stars dominate the observations in spectroscopic surveys, synthetic spectra derived from the stellar atmospheric
model and atomic line lists are needed to isolate the DIB signal from the stellar components. However, the inappropriate modeling
of stellar lines close to the DIB signal could introduce additional uncertainties in DIB measurements. Moreover, when the stellar 
residuals are comparable to the DIB features in terms of strength, a pseudo-fitting is hard to distinguish and could lead to a bias 
in the measurements of DIB parameters. To overcome this limitation, \citet{Kos2013} developed a data-driven method, called the
``best neighbor method'' (BNM), to build artificial stellar templates for the observed spectra in the vicinity of the DIB feature
(DIB window). Specifically, BNM first separates the whole spectroscopic sample into a target sample (spectra containing DIB signals)
and a reference sample (spectra without DIB signals). The reference sample is usually constituted by sources at high latitudes and 
with low dust extinctions according to the assumption that a small extinction represents a low abundance of ISM species. Then, for 
a given target spectrum, BNM finds its best-matched reference spectra based on a  pixel-by-pixel comparison for the spectral region 
outside the DIB window. Finally, a number of best-matched reference spectra (up to 25 in \citealt{Kos2013}) are averaged to create 
a stellar template for the DIB window. The ISM spectrum within the DIB window, where the DIB signal is detected and measured, is 
defined as the target spectrum divided by the generated stellar template. BNM has been applied to measure DIBs in the spectra from 
RAVE \citep{Kos2013}, Gaia RVS (\citealt{hz2022}; \citealt{Schultheis2023FPR}, hereafter GFPR), and GALAH \citep{Vogrincic2023}. 
Other kinds of data-driven methods have been applied to DIB detection as well. \citet[][hereafter AS23]{Saydjari2023} decomposed 
and recognized stellar components and DIBs in public Gaia RVS spectra with a data-driven prior consisting of $\sim$40\,000 RVS 
low-extinction spectra. \citet{McKinnon2023} built 2nd-order polynomial models of normalized flux as a function of stellar 
parameters for $\sim$17\,000 red clump stars observed in APOGEE and found 84 possible DIBs (25 identified with a confidence level
of 95\% and 10 of them were previously known) in the residuals between observed and modeled APOGEE spectra. 

The third data release of Gaia \citep{Vallenari2023} contains a large number of measurements for DIB\,$\lambda$8621 in about 500\,000 
RVS spectra of individual stars \citep[][hereafter GDR3]{Schultheis2023DR3}. DIB\,$\lambda$8621 was fitted in the ISM spectra derived 
by the synthetic spectra from the General Stellar Parametrizer from spectroscopy ({\gspspec}) module \citep{Recio-Blanco2023}. After 
removing the cases with bad stellar modelings and bad DIB parameters, we defined a high-quality (HQ) DIB sample containing $\sim$140\,000 
sightlines (see Sect. 3 in \citetalias{Schultheis2023DR3} for details). However, \citetalias{Saydjari2023} reported a dependence 
between the fitted central wavelength and the Gaussian width of DIB\,$\lambda$8621 (see their Fig. 1) and attributed these biases 
to the residuals of stellar lines in the vicinity of DIB\,$\lambda$8621 (e.g. \ion{Fe}{i} lines at 8620.51\,{\AA} and 8623.97\,{\AA} 
in vacuum wavelength determined by \citealt{Contursi2021}). In this work, we improve BNM to a machine-learning (ML) approach, that 
replaces the pixel-by-pixel comparison of spectral flux in finding the best-matched reference spectra by ML training. The ML approach 
can directly predict stellar components in the DIB window for the target spectra rather than comparing them with reference spectra 
one by one. Thus, the ML approach can speed up the process and ignore irrelevant features. Furthermore, \citet{Kos2013} down-weighted 
the \ion{Ca}{ii} regions in the comparison between target and reference spectra because \ion{Ca}{ii} lines are too strong to overwhelm 
other stellar features, but the ML model does not need to adjust the weights of stellar lines. We applied the improved BNM to process 
780 thousand RVS spectra published in Gaia DR3 and measured DIB\,$\lambda$8621 as well as the broad DIB around 8648\,{\AA} 
\citep[$\lambda$8648;][]{hz2022}. We did some statistical analysis of the properties of these two DIBs based on a selected reliable 
sample. We compared our new measurements of DIB\,$\lambda$8621 to those in \citetalias{Schultheis2023DR3} and \citetalias{Saydjari2023} 
to estimate the degree of biases of DIB parameters in \citetalias{Schultheis2023DR3}. We note that BNM was already applied to RVS 
spectra in the new Focused Product Release (FPR) of Gaia to measure DIBs $\lambda$8621 and $\lambda$8648 (see \citetalias{Schultheis2023FPR} 
for detailed results), but FPR only contained DIB measurements in stacked ISM spectra and this work measured both of DIBs $\lambda$8621 
and $\lambda$8648 in the RVS spectra of individual stars for the first time.

The paper is outlined as follows: The data processing is described in Sect. \ref{sect:data}. Section \ref{sect:method} introduces
our ML model and the fitting of DIBs $\lambda$8621 and $\lambda$8648. In Sect. \ref{sect:results}, we investigate the intensity 
and kinematic properties of DIB\,$\lambda$8621, analyze the detection of DIB\,$\lambda$8648 in individual RVS spectra, and reassess 
the results of $\lambda$8621 in \citetalias{Schultheis2023DR3}. The correlation between DIB\,$\lambda$8621 strength and hydrogen
abundance, as well as the completeness of the DIB catalog, are discussed in Sect. \ref{sect:discuss}. The main conclusions are 
summarized in Sect. \ref{sect:conclusion}.

\begin{figure*}
  \centering
  \includegraphics[width=16cm]{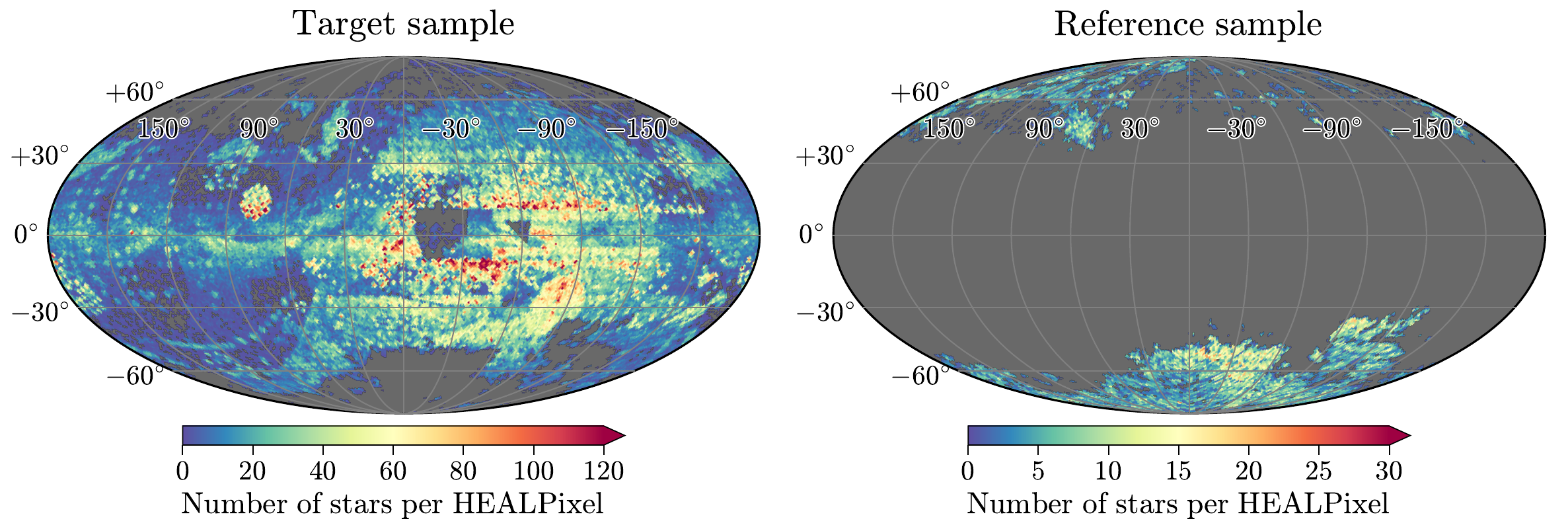}
  \caption{Galactic spatial density distribution of 780\,513 target spectra ({\it left panel}) and 36\,622 reference spectra
  ({\it right panel}), respectively. This HEALPix \citep{Gorski2005} map has a spatial resolution of $0.92^{\circ}$ ($N_{\rm 
  side}\,{=}\,64$).}
  \label{fig:sky}
\end{figure*}

\section{Data Processing} \label{sect:data}

There are 999\,645 RVS spectra ($R\,{\sim}\,$11\,500, mean spectra of epoch observations) published in Gaia DR3 \citep{Vallenari2023} 
which can be accessed
through the Datalink interface of Gaia Archive\footnote{\url{http://cdn.gea.esac.esa.int/Gaia/gdr3/Spectroscopy/rvs_mean_spectrum/}}.
The published RVS spectra were processed by Gaia DPAC Coordination Unit 6 (CU6) and equally resampled between 864 and 870\,nm with 
a spacing of 0.01\,nm (2400 wavelength bins; \citealt{Sartoretti2018,Sartoretti2023}). The spectra were normalized and shifted to 
the rest frame as well. Following the process in \citet{Recio-Blanco2023} and \citetalias{Schultheis2023FPR}, we rebinned RVS spectra 
from 2400 to 800 wavelength pixels, sampled every 0.03\,nm, to increase the S/N. The total wavelength range of RVS spectra used in
this work is between 8471.2 and 8687.5\,{\AA} to ensure no reference spectra containing nan-value fluxes. The DIB window is defined 
as 8600--8680\,{\AA} (267 wavelength pixels). 

After combining with the calibrated distance catalog of \citet{Bailer-Jones2021}, where we made use of their geometric distances, 
there are 996\,900 sources left. We calculated $\EBV$ for these stars by the Planck dust map \citep{Planck2016dust} using the python 
package {\it dustmaps} \citep{Green2018python}. The target sample contains 780\,513 spectra with $\EBV\,{>}\,0.02$\,mag and 
signal-to-noise ratio (S/N) greater than 20. The reference sample contains 36\,622 spectra with $\EBV\,{\leqslant}\,0.02$\,mag, 
$|b|\,{\geqslant}\,30^{\circ}$, and $\rm S/N\,{\geqslant}\,50$. A higher threshold of S/N for the reference sample is for a better 
training set. The density distribution in Galactic coordinates for the target and reference samples is shown in Fig. \ref{fig:sky}. 
\citet{Baron2015a} reported the detection of DIB signals in dust-free regions at high latitudes. These DIB signals in reference 
spectra could introduce an offset or a bias in modeling the DIB profiles in target spectra, but if we assume such DIBs only exist 
in a very small part of the reference sample, the ML model would treat them as irrelevant features and minimize their effect. 
Nevertheless, this problem cannot be quantified due to the lack of DIB maps built with hot stars that are free of the usage of 
reference spectra. 

\section{Method} \label{sect:method}

\subsection{Build stellar templates by a Random Forest model} \label{subsect:RF}

Various supervised learning algorithms could be applied to model the stellar lines in the DIB window. In this work, we build a model
based on the random forest (RF) regression, which is an ensemble bagging method combining a large number of decision trees \citep{Breiman01}. 
The RF model predicts the stellar template within the DIB window (8600--8680\,{\AA}) for a given target spectrum using the part 
outside the DIB window (i.e. 8471.2--8600\,{\AA} and 8680--8687.5\,{\AA}). The model construction and prediction of the RF algorithm
were completed by the python package {\it scikit-learn} \citep{Pedregosa2011}.

\begin{figure*}
  \centering
  \includegraphics[width=16.8cm]{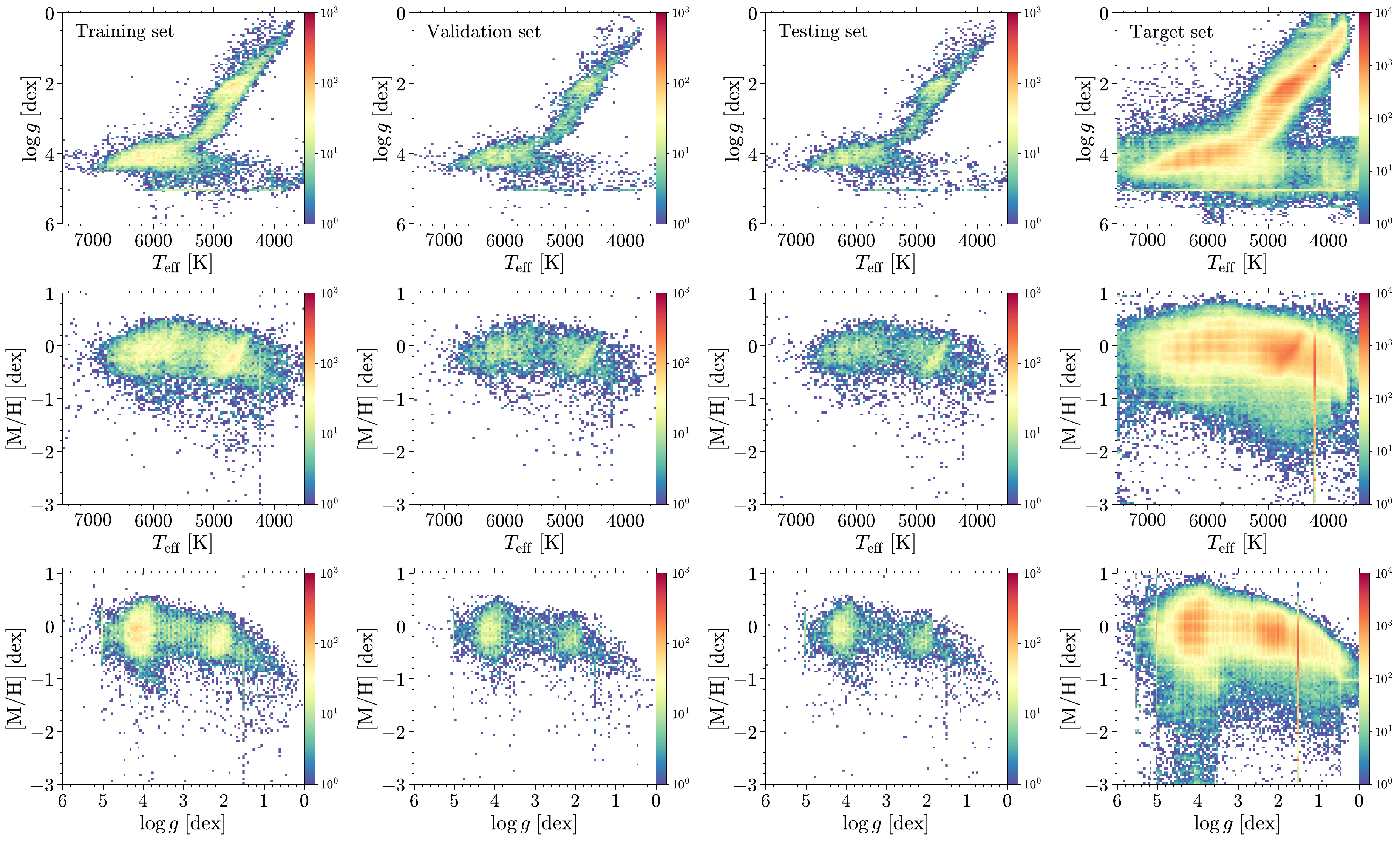}
  \caption{Two-dimensional distributions of stellar atmospheric parameters ($\teff$, $\logg$, $\meta$) from {\gspspec} \citep{Recio-Blanco2023} 
  for the training set, the validation set, the testing set, and the target set, respectively. The color represents the number 
  of stars counted in each bin with a size of $\Delta \teff\,{=}\,40$\,K, $\Delta \logg\,{=}\,0.05$, and $\Delta \meta\,{=}\,0.04$\,dex.}
  \label{fig:ML-params}
\end{figure*}

\begin{figure}
  \centering
  \includegraphics[width=8cm]{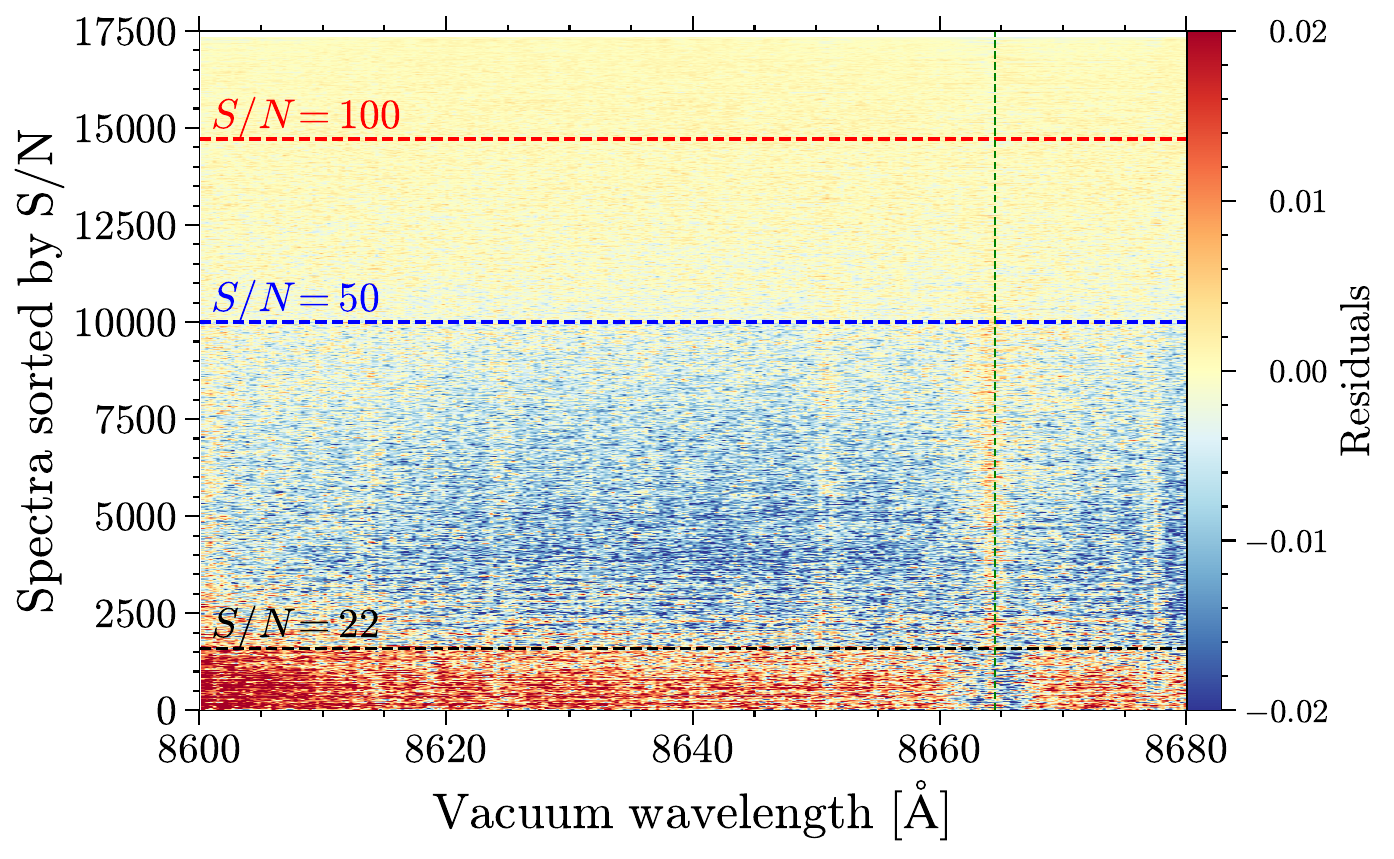}
  \caption{Residuals between the observed and modeled normalized fluxes as a function of the wavelength for 17\,324 RVS spectrum 
  in the testing set. The color represents the residuals and the spectra are sorted by spectral S/N, increasing from bottom to top. 
  $\rm S/N\,{=}\,22$, 50, 100 are indicated by the dashed black, blue, and red lines, respectively. The \ion{Ca}{ii} line within 
  the DIB window is marked as a dashed green line. }
  \label{fig:test-set}
\end{figure}

\begin{figure}
  \centering
  \includegraphics[width=8cm]{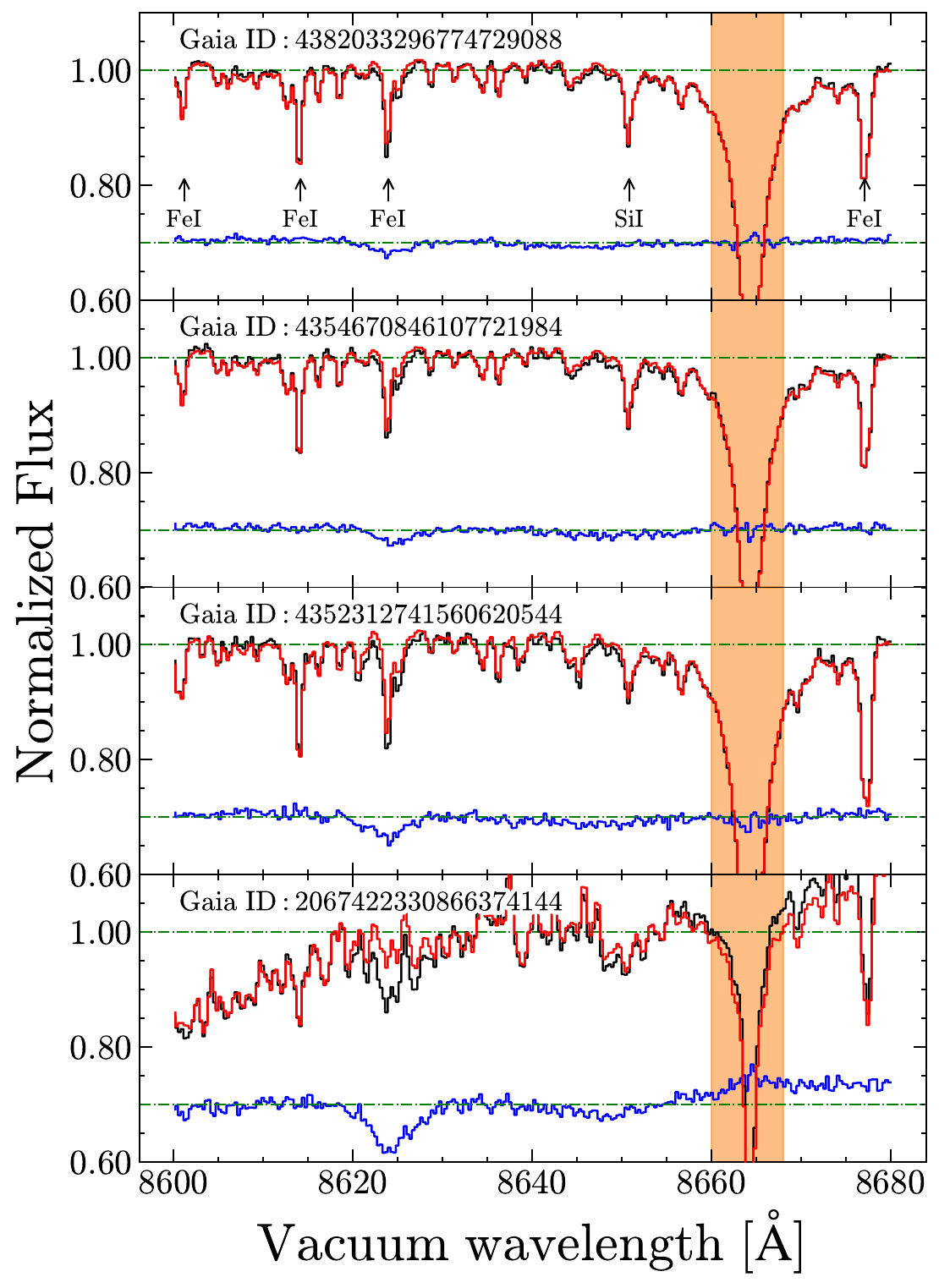}
  \caption{Four examples of the RF prediction within the DIB window. The black and red lines are the observed RVS spectra and the
  predicted stellar parameters, respectively. The blue line is the derived ISM spectrum with an offset of 0.3 for the normalized 
  flux. Orange marks the masked region during the fittings. The Gaia source ID of these targets is indicated. Some typical stellar 
  lines within the DIB window determined by \citet{Contursi2021} are marked as well. The DIB fitting to these ISM spectra is shown 
  in Fig. \ref{fig:fit-example}.}
  \label{fig:rf-example}
\end{figure}

The reference sample was separated into three data sets: the training set containing 21\,974 spectra (60\%), the validation 
set containing 7324 spectra (20\%), and the testing set containing 7324 spectra (20\%). The distributions of the stellar atmospheric
parameters ($\teff$, $\logg$, $\meta$) from {\gspspec} \citep{Recio-Blanco2023} of these sets, as well as the target set, are 
presented in Fig. \ref{fig:ML-params}. The reference sample (training, validation, and testing sets) has a similar coverage 
of stellar parameter space compared to the target sample, mainly covering the main sequence, sub-giant and red-giant branches, 
and a $\meta$ region between --1 and 0.5\,dex. Metal-poor and extremely hot or cool stars are notably missing, but they only form 
a small fraction of the target sample. The stellar parameters are only used to present the space coverage but were not used in our 
RF model. They are also not necessary for BNM but were always used to speed up the BNM process (e.g. \citealt{Kos2013}; \citealt{hz2022}; 
\citetalias{Schultheis2023FPR}). 

Two of the most important parameters for the RF model are the number of trees in the forest (`n\_estimators', NE) and the maximum 
depth of the tree (`max\_depth', MD). Therefore, we trained RF models using various NE ($\{20,40,60,80,100\}$) and 
MD ($\{2,5,10,30,50,100\}$), keeping other parameters as default values in {\it scikit-learn}, and completed the model selection 
by the performance of the validation set. For each pair of NE and MD, we applied the trained RF model to predict the stellar
components in the DIB window for each RVS spectrum in the validation set and calculated the residuals between observed and 
modeled normalized fluxes at each wavelength pixel. The mean residuals of these RVS spectra as a function of the spectral 
wavelength for different pairs of NE and MD are shown in Fig. \ref{appfig:RF-model1}, as well as their standard deviations. For 
small NE and MD, structural residuals can be seen around the \ion{Ca}{ii} line at 8664.5\,{\AA} \citep{Contursi2021}, which means 
a bad modeling of this strong line. With the increase of NE and MD, the \ion{Ca}{ii} is better modeled and the standard deviation 
also becomes smaller. Further, the degree of dispersion of the residuals becomes similar for $\rm NE\,{\geqslant}\,60$ and $\rm 
MD\,{\geqslant}\,30$. Meanwhile, their mean residuals are slightly smaller than zero, which could be caused by the imperfect 
normalization of the observed spectra. Some mean residuals with small NE and MD are very close to zero, but apparently, their 
dispersion is more significant. Some weak structural residuals exist even for the maximum NE and MD, but they are only at the order 
of $10^{-4}$. On the other hand, Fig. \ref{appfig:RF-model2} shows the mean of the absolute residuals (MAR), taken along the 
wavelength within the DIB window, of each RVS spectra in the validation set as a function of the spectral S/N. MAR decreases 
with the increase of S/N presenting a strong dependence. The dependence breaks for $\rm S/N\,{\gtrsim}\,300$, where the dispersion 
of MAR represents the robustness of the RF model for different types of spectra. And MAR would dramatically increase for the stars 
with extreme parameters. For $\rm NE\,{\geqslant}\,20$ and $\rm MD\,{\geqslant}\,30$, the distribution of MAR becomes similar and 
MAR is smaller than 0.01 for $\rm S/N\,{\gtrsim}\,100$. Because of the similar performance of the validation set for large 
NE and MD, we selected a final parameter pair of $\rm NE\,{=}\,100$ and $\rm MD\,{=}\,50$. We also trained RF models with larger 
NE and MD and found that the performance improvement was not significant. For $\rm NE\,{=}\,200$ and $\rm MD\,{=}\,300$, the 
mean and the standard deviation of the residuals is ${-}0.46$ and 117.58, which is only slightly better than those for $\rm 
NE\,{=}\,100$ and $\rm MD\,{=}\,50$.

With selected $\rm NE\,{=}\,100$ and $\rm MD\,{=}\,50$, we applied the RF model to the testing set to evaluate its performance on 
the target sample. Because 62\% of the target spectra have a S/N lower than 50, we randomly selected 10\,000 reference spectra with 
$\rm 20\,{\leqslant}\,S/N\,{<}\,50$ and added them to the testing set (a total of 17\,324 reference spectra). Figure \ref{fig:test-set} 
presents the residuals as a function of wavelength for each spectrum in the testing set, sorted by the spectral S/N. For $\rm 
S/N\,{>}\,50$, the distribution of residuals is generally uniform along the wavelength, which indicates good modeling for the 
stellar lines. Only the residuals near the \ion{Ca}{ii} line (indicated by a dashed green line in Fig. \ref{fig:test-set}) are 
more significant than its vicinity. The performance of the RF model becomes worse for spectra with lower S/N, showing systematic 
differences between the observed spectra and the modeled stellar template and structural residuals near the stellar lines. The 
systematic difference could be caused by the imperfect normalization for low-S/N RVS spectra. We applied a linear continuum in the 
DIB model (see Sect. \ref{subsect:dib-fit}) that could reduce such effect. 

Figure \ref{fig:rf-example} shows the stellar templates predicted by the RF model for four RVS spectra within the DIB window.
Strong stellar lines, such as \ion{Fe}{i} and \ion{Si}{i}, are well modeled, and the feature of $\lambda$8621 can be clearly seen
in the derived ISM spectra. The residuals near the center of \ion{Ca}{ii} slightly increase. In the last example (bottom), the RVS
spectrum was not perfectly normalized, but the RF model can still properly predict the stellar components. Further, the ISM spectrum 
can be well-fitted by our DIB model with a linear continuum (see the bottom panel in Fig. \ref{fig:fit-example}). 

\subsection{Fit DIBs in ISM spectra} \label{subsect:dib-fit}

With the trained RF model, ISM spectra can be obtained by the modeled stellar templates in the DIB window divided by the observed
spectra for each RVS source in the target sample. The S/N of the ISM spectra is calculated between 8602 and 8612\,{\AA} as $\rm 
mean(flux)/std(flux)$. Following our previous works (\citealt{hz2022}; \citetalias{Schultheis2023FPR}), we model the profiles of 
the two DIBs in the ISM spectra by a Gaussian function (Eq. \ref{eq:gauss}) for $\lambda$8621, a Lorentzian function (Eq. 
\ref{eq:lorentz}) for $\lambda$8648, and a linear function for the continuum (Eq. \ref{eq:cont}):

\begin{equation} \label{eq:gauss}
    G(\lambda;\dibdepth,\diblambda,\dibwidth) = -\dibdepth \times {\rm exp}\left(-\frac{(\lambda-\diblambda)^2}{2\dibwidth^2}\right),
\end{equation}

\begin{equation} \label{eq:lorentz}
    L(\lambda;\dibdepth,\diblambda,\dibwidth)= \frac{-(\dibdepth \dibwidth^2)}{(\lambda-\diblambda)^2+\dibwidth^2},
\end{equation}

\begin{equation} \label{eq:cont}
    C(\lambda;a_0,a_1)= a_0 \times \lambda + a_1.
\end{equation}

\begin{figure}
  \centering
  \includegraphics[width=8cm]{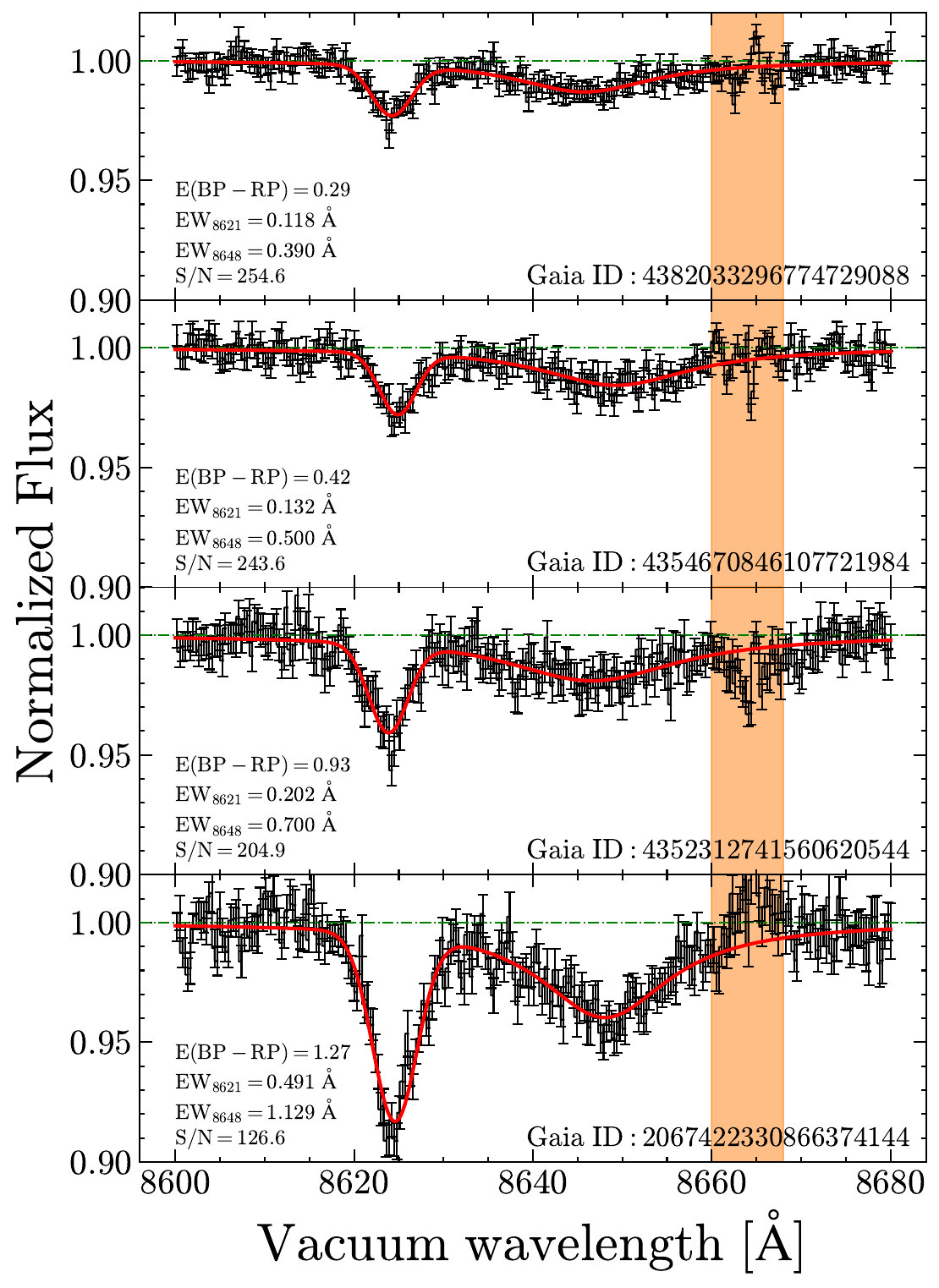}
  \caption{Examples of the fits to DIBs $\lambda$8621 and $\lambda$8648 in four ISM spectra. The black and red lines are the ISM 
  spectra and fitted DIB profiles, respectively, normalized by the fitted linear continuum. The error bars indicate the observational
  flux uncertainties of the RVS spectra. Orange marks the masked region during the fittings. The Gaia source ID of these targets,
  $\EBPRP$ from \citet{Andrae2023}, the EWs of the two DIBs ($\rm EW_{8621}$ and $\rm EW_{8648}$), and the S/N of the ISM spectra
  are indicated as well.  }
  \label{fig:fit-example}
\end{figure}

\noindent where $\dibdepth$ and $\dibwidth$ are the depth and width of the DIB profile, $\diblambda$ is the measured central 
wavelength, $a_0$ and $a_1$ describe the linear continuum, and $\lambda$ is the wavelength. Subscripts `8621' and `8648' are used 
below to distinguish the profile parameters of the two DIBs. The full parameters of the DIB model are $\Theta=\{\mathcal{D}_{8621},
\lambda_{8621},\sigma_{8621},\mathcal{D}_{8648},\lambda_{8648},\sigma_{8648},a_0,a_1\}$. A Markov chain Monte Carlo (MCMC) 
procedure \citep{Foreman-Mackey13} was performed to implement the parameters optimization with flat and independent priors for the 
DIB parameters. The best estimates of the DIB parameters and their statistical uncertainties were taken in terms of the 50th, 16th, 
and 84th percentiles of the posterior distribution drawn by the MCMC procedure. We refer to Sect. 3.3 in \citetalias{Schultheis2023FPR} 
for a very detailed description of the DIB model, priors, and the MCMC fitting procedure. We kept using the masked region between 
8660 and 8668\,{\AA} during the fitting, although the RF algorithm modeled the \ion{Ca}{ii} line much better than BNM. We note that
the uncertainties of the ISM spectra used in the MCMC fitting include only the observational flux errors of the RVS spectra because
the RF model cannot estimate the uncertainty of their predictions, so the total uncertainties would be underestimated. According to 
Eqs. \ref{eq:gauss} and \ref{eq:lorentz}, the equivalent width (EW), representing the DIB strength, for $\lambda$8621 is calculated 
as $\rm EW_{8621} =\sqrt{2\pi} \times \mathcal{D}_{8621} \times \sigma_{8621}$ and for $\lambda$8648 as $\rm EW_{8648}=\pi \times 
\mathcal{D}_{8648} \times \sigma_{8648}$. The lower (16\%) and upper (84\%) confidence levels of EW were estimated by the distributions 
of $\dibdepth$ and $\dibwidth$ drawn from the MCMC posterior samplings. The full width at half maximum (FWHM) of the two DIBs is 
calculated as $\rm FWHM_{8621}=\sqrt{8{\rm ln}(2)} \times \sigma_{8621}$ for the Gaussian profile of $\lambda$862.1 and $\rm 
FWHM_{8648}=2 \times \sigma_{8648}$ for the Lorentzian profile of $\lambda$8648.

Four examples of the DIB fits are shown in Fig. \ref{fig:fit-example}, whose ISM spectra are sorted by $\EBPRP$ calculated by 
\citet{Andrae2023}. $\rm EW_{8621}$ and $\rm EW_{8648}$ both increase with $\EBPRP$. The profile of DIB\,$\lambda$8621 is prominent 
in all the ISM spectra while the DIB\,$\lambda$8648 has a much shallower and broader profile than that of $\lambda$8621. Because of 
the very small $\mathcal{D}_{8648}$, it is much more difficult to measure $\lambda$8648 than $\lambda$8621 in the ISM spectra 
derived from the individual RVS spectra. Additionally, the masked region, where the residual of the \ion{Ca}{ii} line is clear, 
also affects the fit to the red wing of the $\lambda$8648 profile. 

We did an injection test following the principles in \citetalias{Saydjari2023}. The details and discussions are presented in Appendix 
\ref{appsect:intest}. In summary, the distribution of the Z-scores as a function of the injected DIB parameters and the S/N of the 
ISM spectra is perfectly uniform, which is highly consistent with the findings in \citetalias{Saydjari2023}, primarily validating 
our RF model and the DIB fittings.

\section{Results} \label{sect:results}


\begin{figure*}
  \centering
  \includegraphics[width=16.8cm]{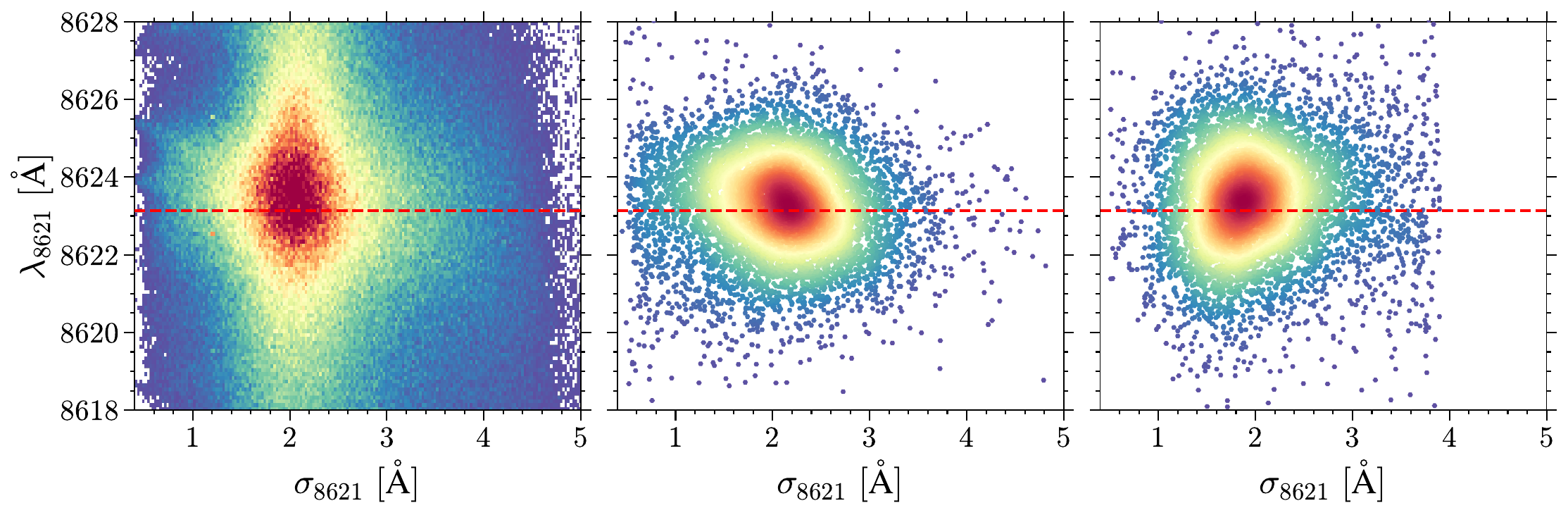}
  \caption{The number density of the measured DIB\,$\lambda$8621 as a function of the Gaussian width ($\sigma_{8621}$) and the central
  wavelength ($\lambda_{8621}$) of the fitted DIB profile for the full target sample (left), 8388 selected DIB measurements (middle), 
  and the reliable measurements in \citetalias{Saydjari2023} (right). The color of the left panel represents the number of DIBs calculated
  in 0.025\,{\AA}$\times$0.1\,{\AA} bins. In the middle and right panels, the color represents the number density estimated by a 
  Gaussian KDE. The dashed red line indicates the rest-frame wavelength of DIB\,$\lambda$8621 determined in this work (8623.141\,{\AA},
  see Sect. \ref{subsect:kinematics}).}
  \label{fig:lambda-sigma}
\end{figure*}

\subsection{Select a reliable DIB catalog} \label{subsect:catalog}

The DIB fitting was done for 780\,513 ISM spectra derived from the target sample. To select reliable measurements, we calculated 
the $\chi^2_{\rm dof}=\chi^2_{\rm tot}/{\rm dof}$, where $\chi^2_{\rm tot}$ is the total $\chi^2$ between the fitted DIB profile 
and the ISM spectrum and $\rm dof\,{=}\,259$ is the degree of freedom of the DIB model (267 wavelength pixels and 8 DIB parameters).  
We applied a cut on $0.71\,{<}\,\chi^2_{\rm dof}\,{<}\,1.41$. The borders were applied by \citetalias{Saydjari2023} based on their
injection test. Generally, stricter borders will get more accurate measurements to some extent but will lose more cases as well. 
We tried other borders and found that 0.71 and 1.41 were proper borders to exclude many of the noisy cases with pseudo fittings.
We further required $\mathcal{D}_{8621}\,{>}\,3R_C$, where $R_C$ is defined as the standard deviation of the residuals between
the flux of the ISM spectrum and the fitted DIB profile in a range from $\diblambda\,{-}\,3\dibwidth$ to $\diblambda\,{+}\,3\dibwidth$.
$R_C$ represented the noise level close to the center of the DIB profile, and $3R_C$ was a strict cut for the strong DIB\,$\lambda$8621.
While for $\lambda$8648, we only required $\mathcal{D}_{8648}\,{>}\,R_C$. This criterion cannot ensure a good measurement of 
$\lambda$8648 but can exclude very noisy cases. It will, on the other hand, introduce a selection bias to DIB\,$\lambda$8621 as
the two DIBs are not necessary to exist in the spectra together. Finally, we required the S/N of ISM spectra to be greater than 20. 
All these cuts left us with 8388 DIB measurements. 

Figure \ref{fig:lambda-sigma} shows the $\dibwidth-\diblambda$ distribution of DIB\,$\lambda$8621 for the full target sample (left
panel), the selected 8388 DIB measurements (middle panel), and the reliable measurements in \citetalias{Saydjari2023}\footnote{The 
catalog can be accessed via \url{https://zenodo.org/record/8303423}.}. The full target sample has a scattered distribution with a 
high-density region (red region in the figure) around the rest-frame wavelength ($\lambda_0\,{=}\,8623.141$\,{\AA}, see Sect. 
\ref{subsect:kinematics}) and the mean Gaussian width ($\sigma_{8621}\,{\approx}\,2$\,{\AA}; \citealt{HL1991}; \citealt{JD1994};
\citetalias{Saydjari2023}; \citetalias{Schultheis2023FPR}) of DIB\,$\lambda$8621. The impact of the stellar line residuals is 
apparent for small $\sigma_{8621}$ that $\lambda_{8621}$ concentrated around the \ion{Fe}{i} lines around 8624 and 8625\,{\AA}. 
The selected DIB measurements in this work and \citetalias{Saydjari2023} both presented a quasi-Gaussian distribution in the 
$\dibwidth-\diblambda$ panel, with some deviations that behave in different ways in the two catalogs. The number density of each 
point in the $\dibwidth-\diblambda$ distribution was estimated by a Gaussian Kernel Density Estimation (KDE) using the python 
package {\it scipy} \citep{scipy}. Our catalog contains a tail toward small $\sigma_{8621}$ ($\lesssim$1\,{\AA}), and $\lambda_{8621}$ 
of these cases seems to be affected by the \ion{Fe}{i} line. This is because we set an initial guess of $\sigma_{8621}\,{=}\,1.2$\,{\AA} 
in the MCMC fitting for all the cases so that noisy ISM spectra with weak DIB features would obtain a fitted $\sigma_{8621}$ around 
this initial guess. On the other hand, \citetalias{Saydjari2023} contains more cases with large $\sigma_{8621}$ ($\gtrsim$3\,{\AA}) 
than our catalog and their $\lambda_{8621}$ there is more scattered as well. The median $\sigma_{8621}$ in our catalog is 2.13\,{\AA} 
which is slightly larger than that in \citetalias{Saydjari2023} (1.92\,{\AA}). This may be due to the different pixel sizes of RVS
spectra used in this work (0.3\,{\AA}\,pixel$^{-1}$)\footnote{We rebinned the RVS spectra following the process in Gaia DR3, see
\citet{Recio-Blanco2023} and \citetalias{Schultheis2023DR3}.} and in \citetalias{Saydjari2023} (0.1\,{\AA}\,pixel$^{-1}$). We 
further applied cuts to constrain $\lambda_{8621}$ between 8620 and 8626\,{\AA} and $\sigma_{8621}$ within 1--4\,{\AA}. Hence, 
the final selected DIB catalog contains 7619 measurements. This DIB catalog can be accessed via the CDS database, Strasbourg, 
France \citep{CDS2000}.

\begin{figure}
\centering
\includegraphics[width=8cm]{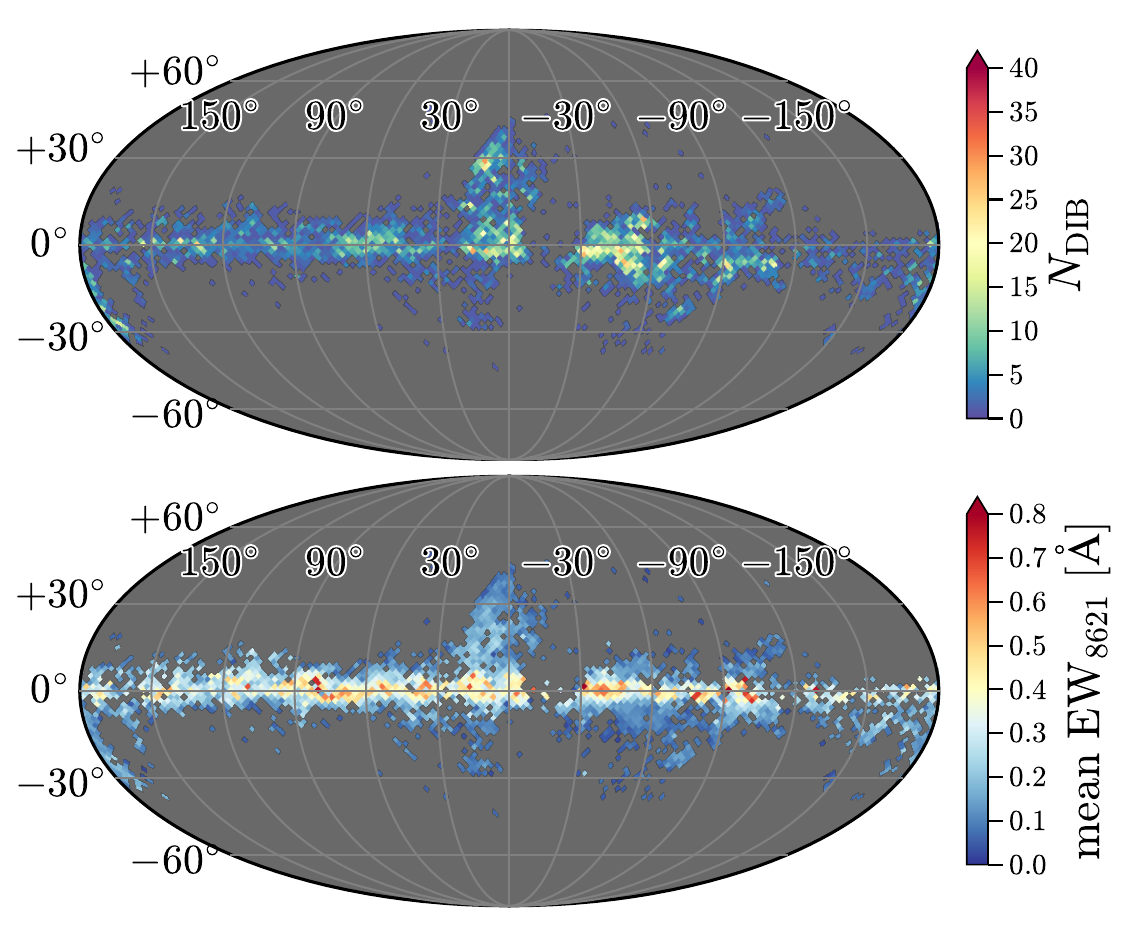}
\caption{The distribution of the number of DIB measurements ($N_{\rm DIB}$, {\it upper panel}) and the mean $\rm EW_{8621}$ 
({\it lower panel}) in a Galactic projection for the selected DIB catalog (7619 measurements). The $N_{\rm DIB}$ and mean $\rm 
EW_{8621}$ are calculated in each HEALPixel with a resolution of $1.83^{\circ}$ ($N_{\rm side}=32$).}
\label{fig:ndib-mew} 
\end{figure}

The reliable DIB catalog in this work was constructed only by simple cuts on $\chi^2$, noise level (S/N and $R_C$), and the DIB 
parameters ($\lambda_{8621}$ and $\sigma_{8621}$). It presents a Gaussian-like $\dibwidth-\diblambda$ distribution without any 
significant impacts of stellar lines and shows a good correlation between $\rm EW_{8621}$ and the dust reddening (see Sect. 
\ref{subsect:dib-dust}). The catalog certainly contains pseudo fittings, such as the outliers seen in the $\dibwidth-\diblambda$ 
diagram, but they should only take a very small part of the catalog after the quality control and only have little impact on the 
statistical analysis of the DIB properties. Furthermore, as DIBs are weak features, investigation of specific fittings would need 
a visual inspection of their ISM spectra. Figure \ref{fig:ndib-mew} shows the Galactic distribution of the number of DIB measurements 
and the mean $\rm EW_{8621}$ in the DIB catalog. The detected DIBs have an uneven distribution and concentrate in the Galactic middle 
plane and some prominent molecular regions, with a remarkable extent to high latitudes in the directions of the Galactic center (GC) 
and anti-center (GAC). Like dust reddening, large mean $\rm EW_{8621}$ focus on the Galactic plane with $|b|\,{\lesssim}\,5^{\circ}$ 
and decreases on average with the increase in latitudes. 

\begin{figure}
\centering
\includegraphics[width=8cm]{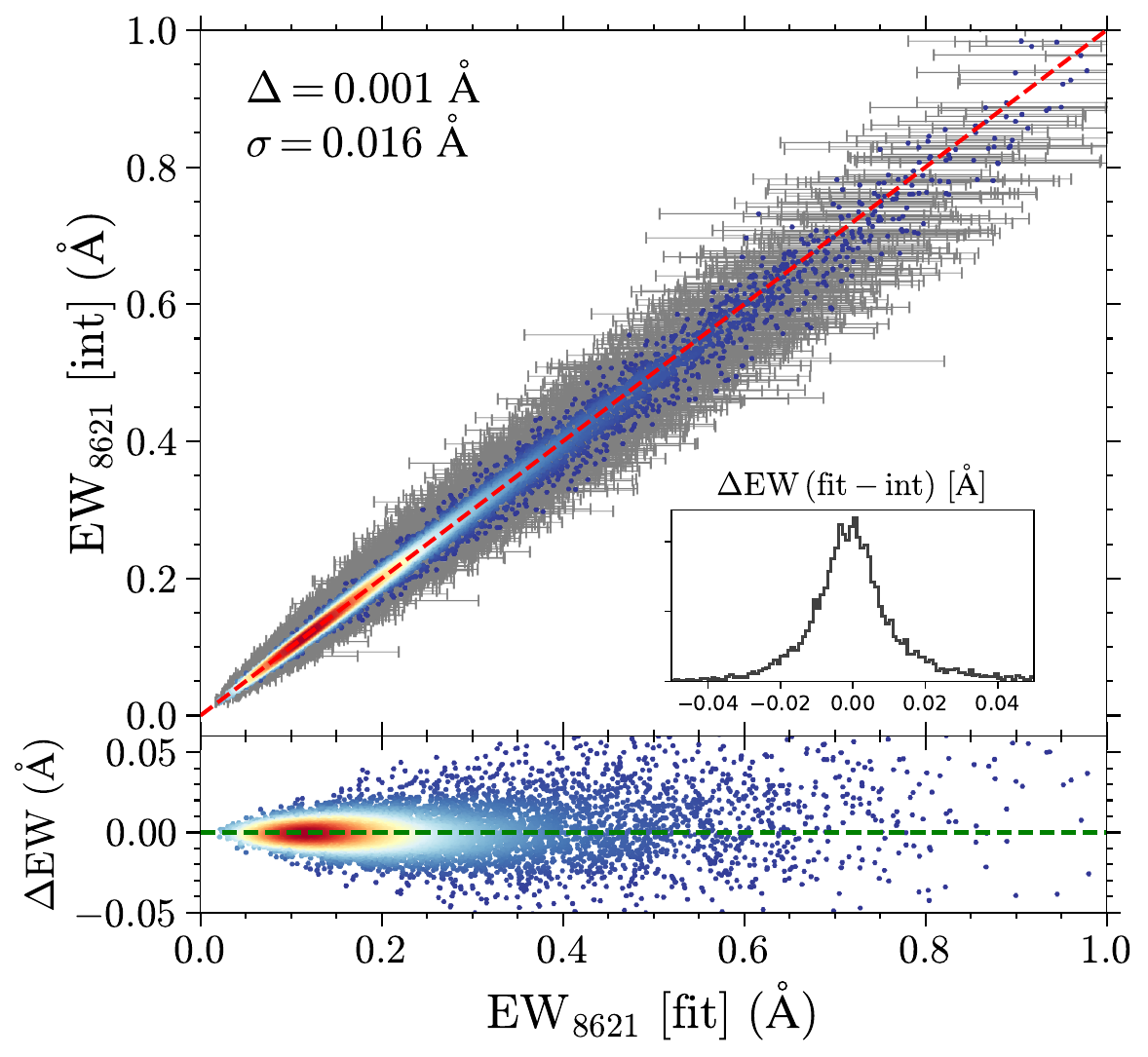}
\caption{{\it Upper panel:} Comparison between the fitted and integrated $\rm EW_{8621}$ for 7619 measurements in the DIB catalog. 
The color represents the number density estimated by a Gaussian KDE. The gray bars show the uncertainty of fitted $\rm EW_{8621}$. 
The dashed red line traces the one-to-one correspondence. A zoom-in panel shows the distribution of the EW difference ($\rm \Delta 
EW=EW_{fit}-EW_{int}$). The mean ($\Delta$) and standard deviation ($\sigma$) of $\rm \Delta EW$ are indicated. {\it Lower panel:} 
Distribution of $\rm \Delta EW$ as a function of fitted $\rm EW_{8621}$.  }
\label{fig:fit-int} 
\end{figure}

To validate the DIB catalog, we compared the fitted and integrated EW of DIB $\lambda$8621 for all the 7619 measurements. Because 
the profiles of $\lambda$8621 and $\lambda$8648 could be overlapped with each other, the fitted profile of $\lambda$8648 was first 
subtracted from the ISM spectra normalized by the fitted linear continuum, and then the rest part within $\lambda_{8621} \pm 
3\sigma_{8621}$ was integrated. Figure \ref{fig:fit-int} shows the comparison between fitted and integrated $\rm EW_{8621}$, as well 
as the EW difference ($\Delta \rm EW = EW_{fit} - EW_{int}$) as a function of fitted $\rm EW_{8621}$. The fitted and integrated $\rm
EW_{8621}$ are highly consistent with each other with a mean difference of only 0.001\,{\AA} and a standard deviation of 0.016\,{\AA}. 
$\Delta \rm EW$ is smaller than the uncertainty of $\rm EW_{8621}$ (0.031\,{\AA} on average) for over 96\% of the measurements, and 
90\% of $\Delta \rm EW$ is smaller than 0.023\,{\AA}. The difference between fitted and integrated $\rm EW_{8621}$ tends to increase 
for the measurements with large $\rm EW_{8621}$. These measurements were done in the ISM spectra with generally lower S/N, where 
the residuals of stellar lines would dramatically increase (the RF model performance becomes worse with low S/N, see Fig. 
\ref{appfig:RF-model2}) and consequently lead to an increase of $\Delta \rm EW$. We checked some ISM spectra with large $\rm EW_{8621}$ 
and $\Delta \rm EW$ and found additional structural features besides the DIB signal. These features are more like from the stellar 
residuals than the possible Doppler splitting caused by multiple ISM clouds along the sightlines. Because they can be far away 
from the center of the DIB profile and usually have a much smaller depth than $\lambda$8621. Despite the heavier influence of the
stellar residuals and noise, the relative EW uncertainty does not increase for large $\rm EW_{8621}$. For $\rm EW_{8621}\,{>}\,0.5$\,{\AA} 
(626 measurements), the fractional error of fitted $\rm EW_{8621}$ is mainly (99.4\%) within 20\% with a mean of 11.9\%, and $\Delta 
\rm EW$ is mainly (99.4\%) smaller than 10\% of fitted $\rm EW_{8621}$. The Doppler broadening caused by the unresolved multiple 
DIB components and the probable intrinsic asymmetry of the DIB profile may contribute to $\Delta \rm EW$ as well. However, the S/N 
of the ISM spectra in this work are not high enough to distinguish these effects from the others.

\subsection{DIB\texorpdfstring{$\,\lambda$}{l}8621 and dust reddening} \label{subsect:dib-dust}

Both DIB EW and dust reddening can be used to map the spatial distribution of ISM species and the Galactic large-scale structures, 
but presently EW generally has a much larger relative uncertainty than reddening. The tight linear correlation between DIB EW and 
dust reddening has been discovered for a set of strong DIBs with early-type stars as background sources \citep[e.g.][]{Munari2008,
Friedman2011,Lan2015} despite the inevitable scatters and outliers \citep[see the review of][]{Krelowski2018}. On the other hand, 
the degree of dispersion between DIB EW and dust reddening usually increases by an order of magnitude for the spectroscopic survey 
data set which is dominated by late-type stars (see e.g. \citealt{Kos2013}; \citealt{Zasowski2015c}; \citetalias{Schultheis2023DR3}). 
The relatively lower S/N of survey spectra (compared to specifically designed DIB observations) and the difficulty in modeling the 
atmospheric components of late-type stars certainly contribute to the increase of dispersion in the DIB--dust correlation. However, 
the numerous observations in the survey should contain some sightlines where the DIB carriers and dust grains are not spatially 
associated with each other because nowadays the dust grain is not considered as a candidate of the DIB carrier \citep{Cox2007,Cox2011}.

\begin{figure}
\centering
\includegraphics[width=8cm]{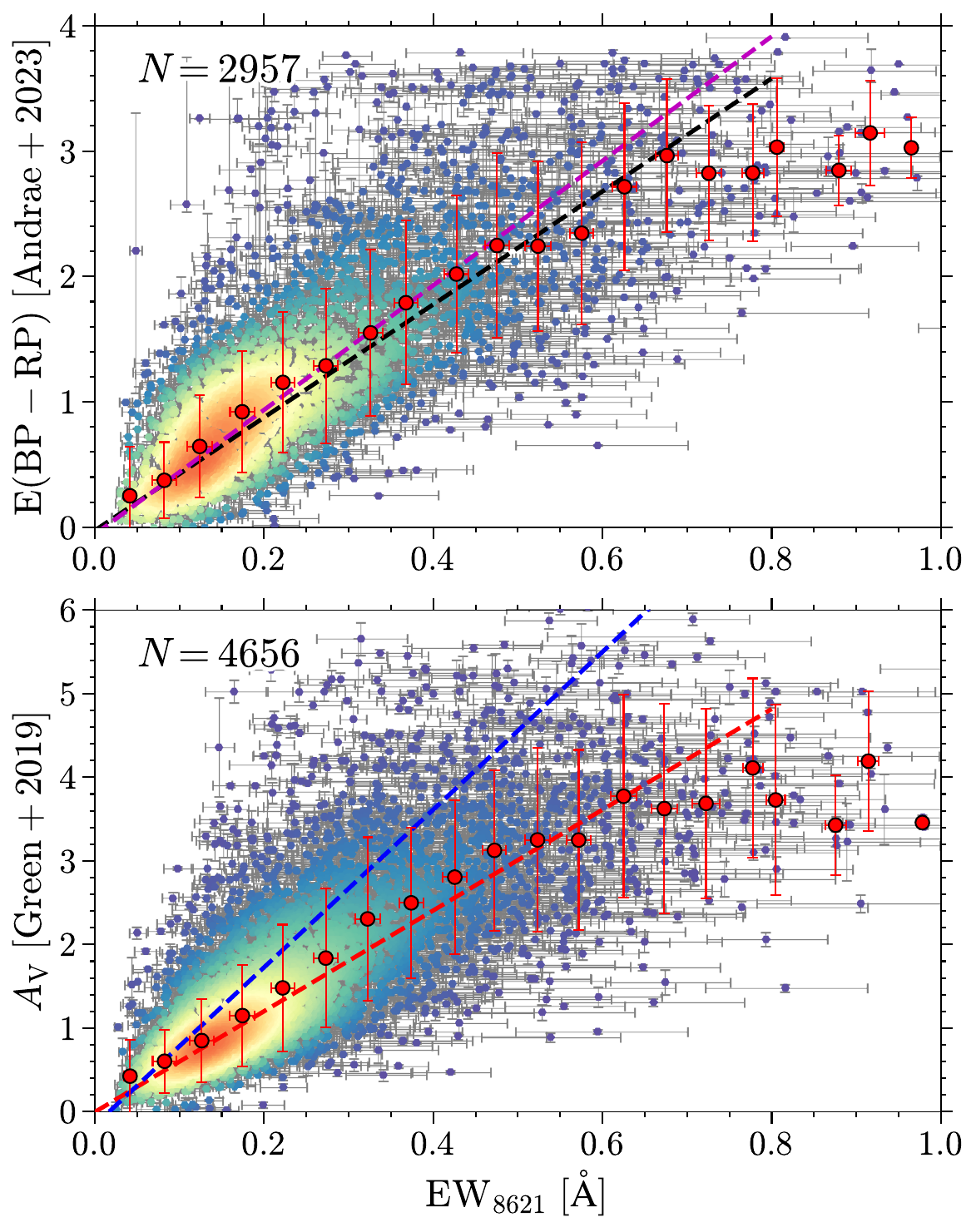}
\caption{Correlation between $\rm EW_{8621}$ and dust reddening for 2957 cases with $\EBPRP$ from \citet{Andrae2023} shown in the
{\it upper panel} and for 4656 cases with $\Av$ from \citet{Green2019} shown in the {\it lower panel}. The color of the scattered
points represents their number density estimated by the Gaussian KDE. The red dots and their color bars are the median values 
and the standard deviations calculated in each $\rm EW_{8621}$ bin with a step of 0.05\,{\AA}. The linear fits to $\rm EW_{8621}$ 
and dust reddening from previous works are overplotted as dashed lines: magenta for \citetalias{Schultheis2023FPR}, black and red 
for \citetalias{Schultheis2023DR3}, and blue for \citetalias{Saydjari2023}. } 
\label{fig:ew-reddening} 
\end{figure}

We reviewed the correlation between DIB\,$\lambda$8621 and dust reddening with our new DIB measurements and two sources of reddening. 
There are 2957 cases in our DIB catalog having $\EBPRP$ from \citet{Andrae2023} and 4656 cases having $\Av$ from \citet{Green2019}. 
The best-estimated $\EBPRP$ and its lower and upper confidence levels for the target stars were accessed via the Gaia Archive. $\Av$
and its uncertainty was obtained by the `bayestar' module in {\it dustmaps} using the `percentile' mode. $\Av$ equals 2.742 times
the reddening unit given by `bayestar' \citep{Green2019}. The scatter plot between $\rm EW_{8621}$ and dust reddening is shown 
in Fig. \ref{fig:ew-reddening} for $\EBPRP$ (upper panel) and $\Av$ (lower panel), respectively, with the median values and standard
deviations taken in each $\rm EW_{8621}$ bin with a step of 0.05\,{\AA} (red dots). The median dots present a good linear relationship
between $\rm EW_{8621}$ and dust reddening for both $\EBPRP$ and $\Av$ for $\rm EW_{8621}\,{\lesssim}\,0.5$\,{\AA}, with a 
deviation at larger $\rm EW_{8621}$. \citetalias{Saydjari2023} reported larger $\Av$ than expected when the median dots deviated
from their linear fit to $\rm EW_{8621}$ and $\Av$, but the median $\Av$ oppositely becomes smaller than expected in our work in 
such regions. This is only because the median dots were taken in $\rm EW_{8621}$ bins in this work but in $\Av$ bins in 
\citetalias{Saydjari2023}. We checked a part of the outliers, for example with large reddening but small $\rm EW_{8621}$, and 
found that many of the DIB measurements have proper DIB parameters and their ISM spectra clearly contain DIB signals by a visual 
inspection. This verifies that the DIB and dust are not necessary to appear together. They could statistically present a linear 
relationship only due to the accumulation of different ISM species along the sightline. Therefore, DIB EW may not be a good 
proxy for dust reddening in specific directions, and their ratio varies with the investigated samples as well. 

The linear fits to $\rm EW_{8621}$ and dust reddening done in previous works are also plotted as dashed lines in Fig. \ref{fig:ew-reddening}. 
For $\EBPRP$, the tendency of the median dots is consistent with the fitted line of \citetalias{Schultheis2023DR3} (black) and of 
\citetalias{Schultheis2023FPR} (magenta). Further, the standard deviation of the individual measurements in each $\rm EW_{8621}$ 
bin is much larger than the difference between \citetalias{Schultheis2023DR3} and \citetalias{Schultheis2023FPR}. For $\Av$, the 
median dots are closer to the line of \citetalias{Schultheis2023DR3} (red) than that of \citetalias{Saydjari2023} (blue). 
\citetalias{Saydjari2023} got a fitted ${\rm EW_{8621}}/\Av\,{=}\,0.106\,{\pm}\,0.017$\,{\AA}\,$\rm mag^{-1}$, corresponding to 
3.448\,mag\,{\AA}$^{-1}$ of $\EBV/{\rm EW_{8621}}$ which is 57\% larger than the value fitted in \citetalias{Schultheis2023DR3} 
(2.198\,mag\,{\AA}$^{-1}$). This difference is mainly caused by the systematic difference in $\rm EW_{8621}$ (see Fig. \ref{fig:GDR3}) 
but not the control of the bias and uncertainties argued by \citetalias{Saydjari2023}. The $\EBV/{\rm EW_{8621}}$ ratio derived 
in different works \citep[e.g.][]{Wallerstein2007,Munari2008,Kos2013,Puspitarini2015} has a 20\% difference on average 
(see Table 3 in \citetalias{Schultheis2023DR3}). The result of \citetalias{Saydjari2023} is similar to that in \citet{hz2021b} 
which used the Gaia--ESO \citep{Gilmore2012} data set and $\EBV$ from \citet{SFD}. We emphasize that the mean correlations between 
$\rm EW_{8621}$ and dust reddening derived in our series works using the Gaia RVS data (\citetalias{Schultheis2023DR3}, \citealt{hz2022}, 
\citetalias{Schultheis2023FPR}) are consistent with each other within 10\% (see also the discussions in Sect. 5.2 in 
\citetalias{Schultheis2023FPR}). 

\subsection{Kinematics of DIB\texorpdfstring{$\,\lambda$}{l}8621} \label{subsect:kinematics}

\begin{figure}
\centering
\includegraphics[width=8cm]{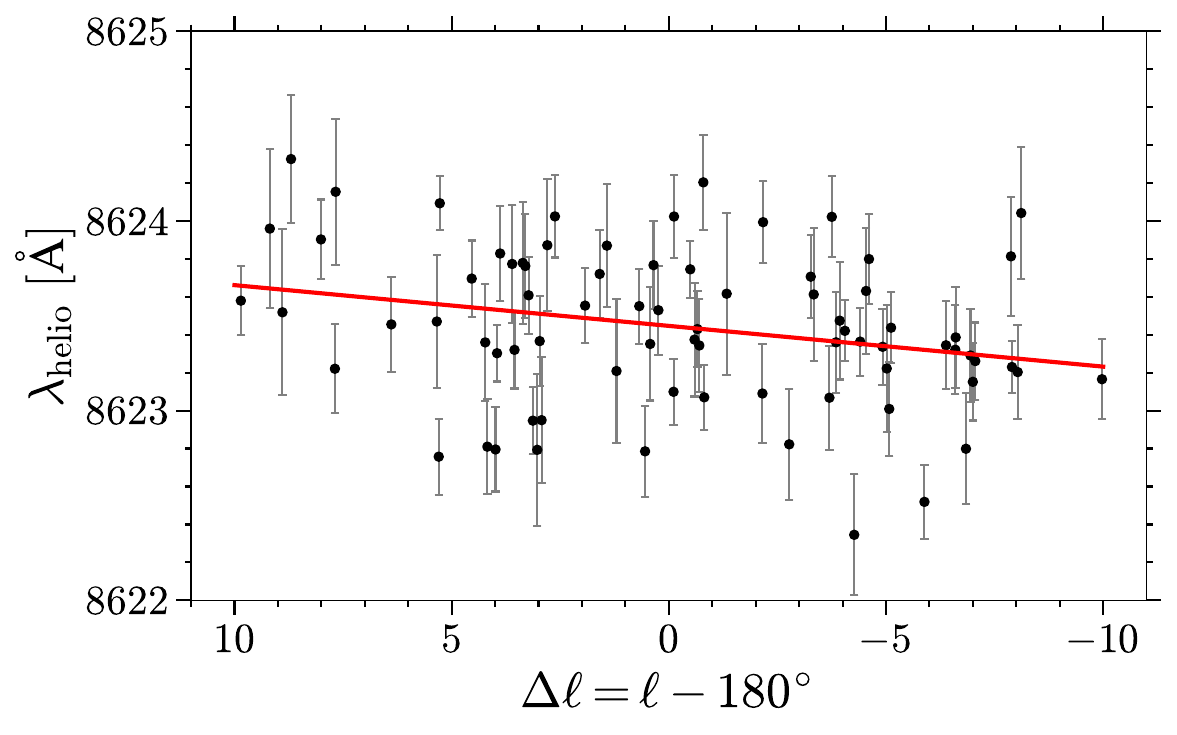}
\caption{Observed central wavelengths in vacuum of DIB\,$\lambda$8621 in the heliocentric frame ($\lambda_{\rm helio}$) as a function 
of the angular distance in longitude from the Galactic anti-center ($\Delta \ell$) for 77 DIB measurements in this work. The black
dots are the individual measurements with the fitted uncertainties. The red line is the linear fit to the black dots.} 
\label{fig:lambda0} 
\end{figure}

To study the kinematics of the carrier of DIB\,$\lambda$8621, the most fundamental and important thing is to determine its rest-frame 
wavelength ($\lambda_0$) which is also necessary to identify the nature of $\lambda$8621 through the comparison to the laboratory 
measurements. In this work, we follow the statistical method which assumes that the DIB radial velocity is null toward the GAC in 
a circular rotational obit, and thus the intercept at $\ell\,{=}\,180^{\circ}$ would indicate $\lambda_0$ (see \citealt{Zasowski2015c}; 
\citetalias{Schultheis2023DR3}; \citetalias{Saydjari2023}). 

As the published RVS spectra have been shifted in the stellar frame where stellar lines are at their rest positions but the DIB 
features are additionally shifted, we have to convert $\lambda_{8621}$ into the heliocentric frame ($\lambda_{\rm helio}$) using 
the radial velocity of stars ($\Vstar$) determined in Gaia DR3 \citep{Katz2023}. We selected 77 DIB measurements in our DIB catalog 
with $|b|\,{<}\,5^{\circ}$, $170^{\circ}<{\ell}<190^{\circ}$, $d\,{\leqslant}\,3$\,kpc, ${\rm err}(\lambda_{8621})\,{<}\,0.5$\,{\AA}, 
and ${\rm err}(\Vstar)\,{<}\,5\,\kms$. The information of their background stars, including the Gaia-DR3 source ID, Galactic coordinates, 
apparent $G$ magnitudes, and stellar atmospheric parameters from {\gspspec}, are listed in Table \ref{apptab:lambda0}, as well as 
their $\lambda_{\rm helio}$. We note that some cases seem to be early-type stars without {\gspspec} estimates of their stellar 
parameters. Nevertheless, by visual inspection, our RF model which does not rely on stellar parameters also works well for these 
spectra, despite that the RVS sample is dominated by late-type stars. Figure \ref{fig:lambda0} presents the slightly linear trend 
of $\lambda_{\rm helio}$ around the GAC, reflecting the projection of the Galactocentric rotation of the DIB carrier \citep{Zasowski2015c}. 
Some tiny deviations of $\lambda_{\rm helio}$ from the linear trend can be seen. Besides the fitting uncertainty of $\lambda_{8621}$, 
the turbulent motion in the DIB cloud and the possible physical changes of DIB shapes and positions \citep[see e.g.][]{Galazutdinov2008b,Krelowski2021} 
would also contribute. The applied statistical method could reduce these effects if no strong systematic deviations exist.
A least-square linear fit to $\lambda_{\rm helio}$ and the angular departure from the GAC ($\Delta \ell$) obtained an intercept 
of $8623.446\,{\pm}\,0.030$\,{\AA}. The uncertainty was estimated by a 2000-times Monte Carlo simulation according to the fitted 
uncertainty of $\lambda_{8621}$. We note that the mean error of $\lambda_{8621}$ from the MCMC fitting is 0.256\,{\AA} for the 77 
selected DIB measurements which is larger than the statistical uncertainty by an order of magnitude. A factor of $c/(c+U_{\odot})$ 
was used to correct the effect of solar motion, where $c$ is the speed of light and $U_{\odot}\,{=}\,10.6\,\kms$ \citep{Reid2019} 
is the radial solar motion toward the GC. Finally, we got a $\lambda_0\,{=}\,8623.141\,{\pm}\,0.030$\,{\AA} for DIB\,$\lambda$8621, 
which is perfectly consistent with the result of \citetalias{Saydjari2023} ($8623.14\,{\pm}\,0.087$\,{\AA}) despite we did not 
consider the distance calibration proposed in \citetalias{Saydjari2023}. This value, nevertheless, is smaller than that of 
\citetalias{Schultheis2023DR3} ($8623.23\,{\pm}\,0.019$\,{\AA}) by 3.0$\sigma$ using our uncertainty for $\lambda_0$ (0.030\,{\AA}). 
Our derived $\lambda_0$ corresponds to 8620.766\,{\AA} in air wavelength that is consistent with most of the literature results 
within 2$\sigma$, such as 8620.7\,{\AA} of \citet{Sanner1978}, 8620.75\,{\AA} of \citet{HL1991}, 8620.79\,{\AA} of 
\citet{Galazutdinov2000a}, 8620.7\,{\AA} of \citet[][after the correction of the solar motion]{Munari2008}, and 8620.83\,{\AA} of 
\citet{hz2021b}. 

\begin{figure}
\centering
\includegraphics[width=8cm]{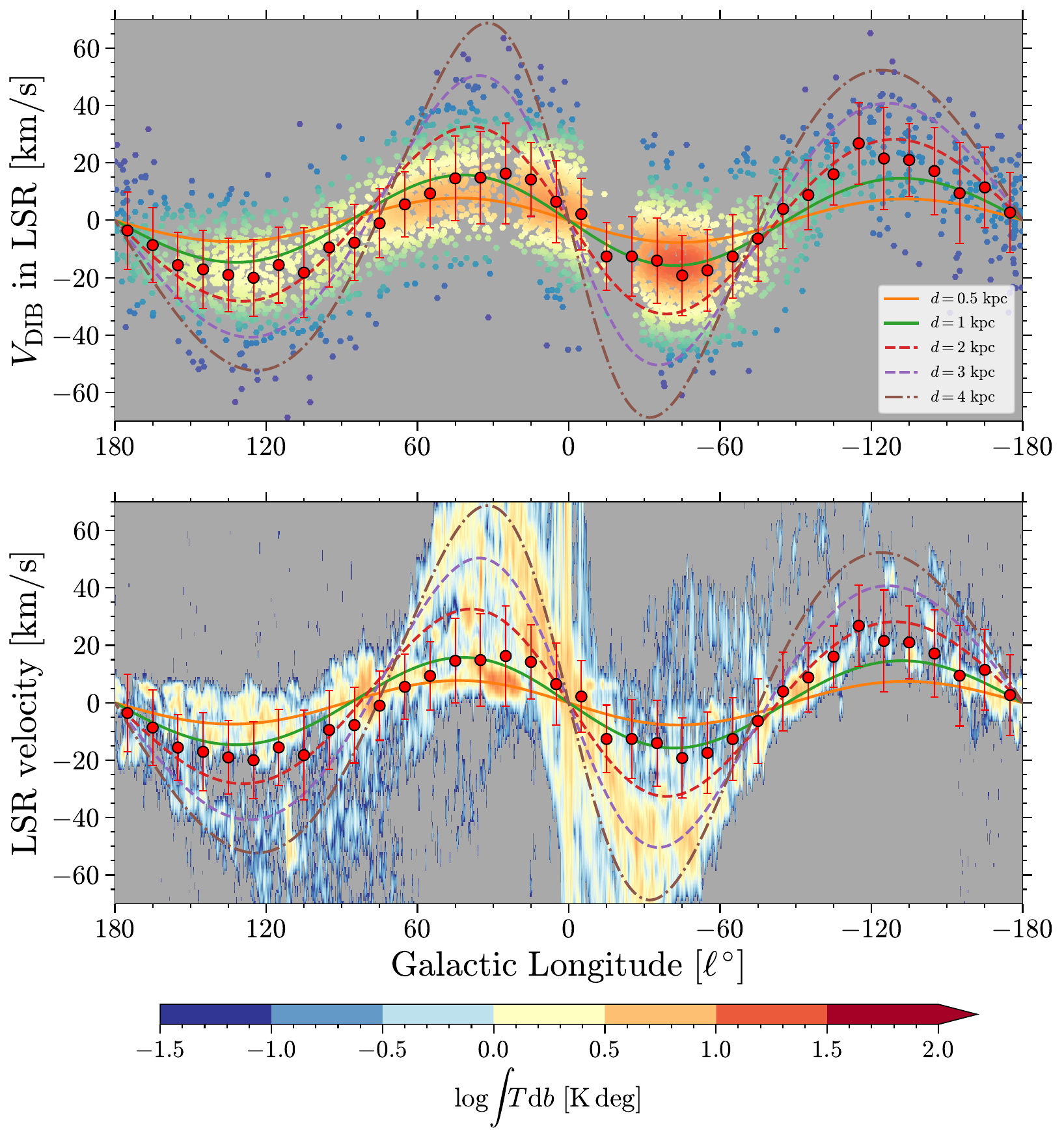}
\caption{{\it Upper panel:}\,$\lambda$8621 carrier ($\Vdib$) along with the Galactic
longitude for 3592 selected DIB measurements. The points are colored by their number density estimated by the Gaussian KDE. The
red dots with error bars are the median $\Vdib$ calculated in each longitude bin with a step of $10^{\circ}$. The colored lines
are theoretical rotation curves calculated with the rotation model in \citet{Reid2019} with different distances from the Sun.
{\it Lower panel:} The median $\Vdib$ are superimposed on the longitude--velocity map of $^{12}$CO $J\,{=}\,(1{-}0)$ emission 
from \citet{Dame2001}. The color scale displays the $^{12}$CO latitude-integrated intensity in a logarithmic scale.} 
\label{fig:RC8621} 
\end{figure}

We applied the same selection criteria to 9763 cases in the HQ DIB catalog in \citetalias{Schultheis2023DR3} measured in the RVS 
spectra that are published in Gaia DR3 and got 67 DIB measurements. The derived $\lambda_0$ is $8623.368\,{\pm}\,0.037$\,{\AA} 
which is even larger than the result of \citetalias{Schultheis2023DR3} with 0.14\,{\AA} (3.8$\sigma$). This result suggests that 
compared to the full Gaia RVS data set, the use of the public sample in DR3 would lead to a red shift of $\lambda_0$. Therefore, 
the smaller $\lambda_0$ determined in this work and in \citetalias{Saydjari2023} is not caused by the selection bias of the sample 
but the systematic difference of $\lambda_{8621}$. As pointed out by \citetalias{Saydjari2023}, $\lambda_{8621}$ in \citetalias{
Schultheis2023DR3} would be affected by the improperly modeled stellar lines. On the other hand, it is not clear if the $\lambda_0$ 
derived in this work and \citetalias{Saydjari2023} with the public RVS sample is also redder than the ``true'' value. We expect 
this problem could be answered by the following analysis of \citetalias{Schultheis2023FPR} or the DIB measurements in Gaia DR4.

With the derived $\lambda_0$, we calculated the radial velocity of the carrier of DIB\,$\lambda$8621 ($\Vdib$) in the Local Standard 
of Rest (LSR)\footnote{The convention between the heliocentric frame and LSR is made with ($U_{\odot},V_{\odot},W_{\odot})\,{=}\,$(
10.6, 10.7, 7.6)\,$\kms$ from \citet{Reid2019}.}. We selected 3592 DIB measurements at low latitudes ($|b|\,{<}\,5^{\circ}$) and with 
accurate $\lambda_{8621}$ (${\rm err}(\lambda_{8621})<0.5$\,{\AA}) and $\Vstar$ (${\rm err}(\Vstar)<5\,\kms$) to present the variation 
of $\Vdib$ along with the Galactic longitude. The rotation of the DIB\,$\lambda$8621 carrier is clearly seen in the upper panel of Fig. 
\ref{fig:RC8621}, overlaid with the theoretical rotation curves with different distances from the Sun. Specifically, for a distance 
from the Sun ($d$), the galactocentric distance is calculated as $R_{\rm GC}=(d^2+R_0^2 -2\,d \cdot R_0 \cdot {\rm cos}(\ell))^{1/2}$,
where $R_0=8.15$\,kpc is the galactocentric distance of the Sun. Then the circular velocity ($\Theta$) is predicted by the Model A5 
in \citet{Reid2019} with $R_{\rm GC}$ and $\ell$, assuming $b=0$. Finally, the radial velocity for a given $d$ and $\ell$ is $V_r = 
(\Theta/R - \Theta_0/R_0) \cdot R_0 \cdot {\rm sin}(\ell)$, where $\Theta_0=236\,\kms$ is the circular velocity of the Sun.

Considering the median $\Vdib$ in each $\Delta \ell\,{=}\,10^{\circ}$ bin (red dots in Fig. \ref{fig:RC8621}), the carrier of 
$\lambda$8621 in the selected sample is mainly located within a kinematic distance of 2\,kpc from the Sun, although the velocities 
from individual DIB measurements are much more scattered. This is a reasonable interpretation by inspecting the mean distances to
the background stars of these DIB signals, which are all larger than 2\,kpc with a minimum of 2.3\,kpc. Moreover, the DIB carriers 
toward the GAC have a larger distance on average than those toward the GC. In the lower panel in Fig. \ref{fig:RC8621}, the median 
$\Vdib$ is compared to the longitude--velocity map of $^{12}$CO $J\,{=}\,(1{-}0)$ emission from \citet{Dame2001}. We made use of 
the momentum-masked cube restricted to a latitude range of ${\pm}5^{\circ}$\footnote{``GOGAL\_deep\_mom.fits.gz'' in 
\url{https://lweb.cfa.harvard.edu/rtdc/CO/CompositeSurveys/}}. The median $\Vdib$ follows the $^{12}$CO velocity curve in the local 
region, especially from $\ell\,{\approx}\,{-}90^{\circ}$ to $\ell\,{\approx}\,{-}180^{\circ}$. The average $\Vdib$ deviation in each 
longitude bin is 13.8\,$\kms$, which prevents the exploration of a finer relationship between DIB\,$\lambda$8621 and $^{12}$CO in 
velocity structures. Nevertheless, more DIBs with large $\Vdib$ in general, can be found in the regions with high-velocity $^{12}$CO 
emission. For instance, the $^{12}$CO velocities between $90^{\circ}$ and $180^{\circ}$ concentrate in two main branches that can 
be interpreted as the Local and Perseus spiral arms \citep[e.g.][]{Reid2019}. Although \citetalias{Saydjari2023} suggested some DIB 
measurements coincide with the $^{12}$CO emission in the Perseus arm, the density distribution of $\Vdib$ there did not bifurcate. 
Such cases consequently only take a very small percentage of the total. For $\ell\,{\approx}\,{-}60^{\circ}$ to $0^{\circ}$, $\Vdib$ 
coincides well with some discrete $^{12}$CO emission at ${-}20$ to ${-}10\,\kms$, suggesting such DIB signals would originate in 
the Local arm.

\subsection{Detection of DIB\texorpdfstring{$\,\lambda$}{l}8648 in individual RVS spectra} \label{subsect:dib8648}

\begin{figure}
\centering
\includegraphics[width=8cm]{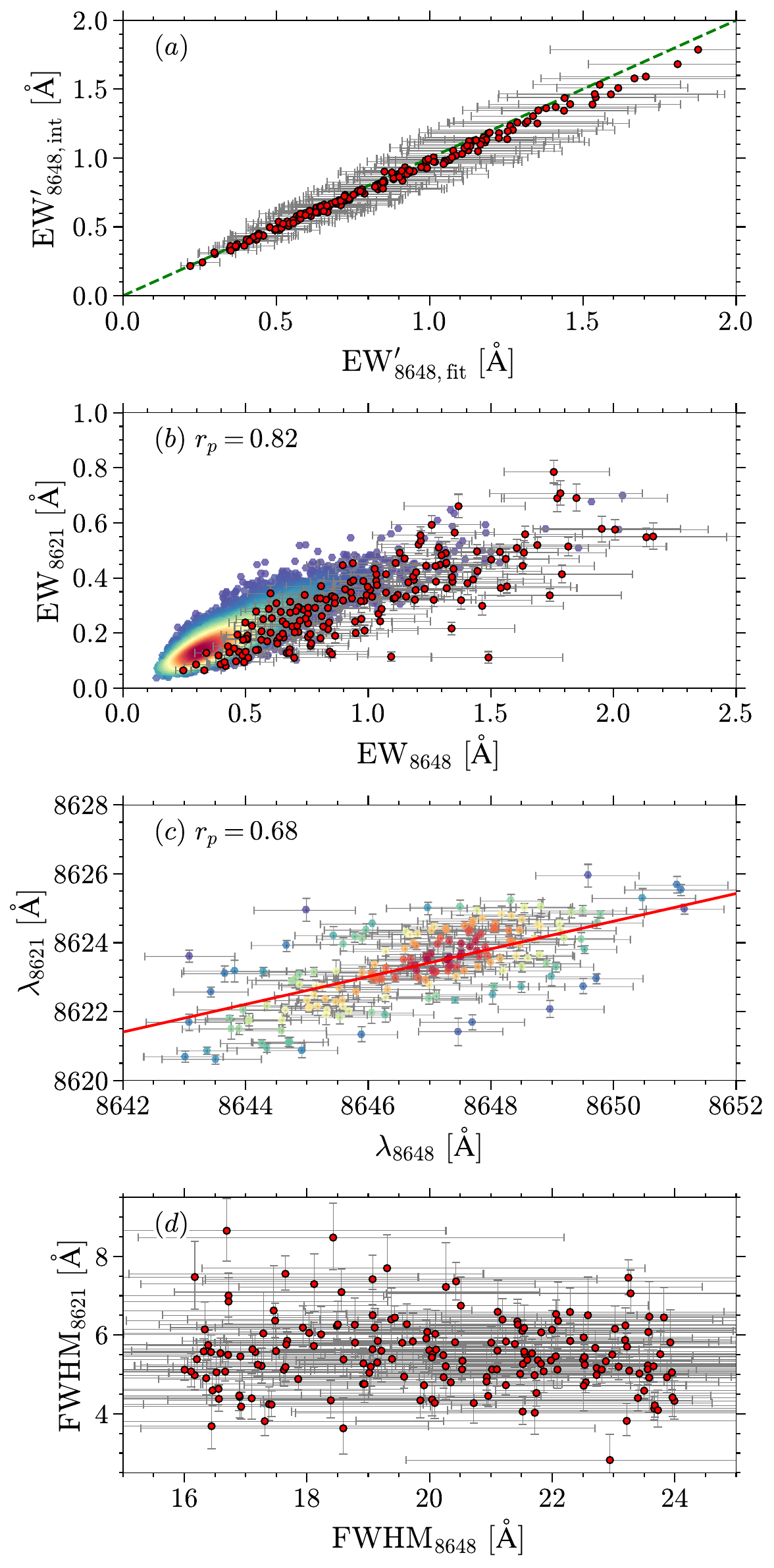}
\caption{Correlations between DIBs $\lambda$8621 and $\lambda$8648 for 179 selected measurements for (a) fitted and integrated 
$\rm EW_{8648}$ outside the masked region between 8660 and 8668\,{\AA}; (b) fitted EW; (c) measured central wavelength; and (d) 
FWHM. The dashed green line in (a) traces the one-to-one correspondence. The colored points in (b) are the results from 
\citetalias{Schultheis2023FPR}. The color in (c) represents the number density estimated by the Gaussian KDE and the red line 
is a linear fit to all the data points. The Pearson coefficient ($r_p$) of the correlation between the parameters of $\lambda$8621 
and $\lambda$8648 for the 179 selected measurements is indicated in (b) and (c).} 
\label{fig:dib8648} 
\end{figure}

\begin{figure}
\centering
\includegraphics[width=8cm]{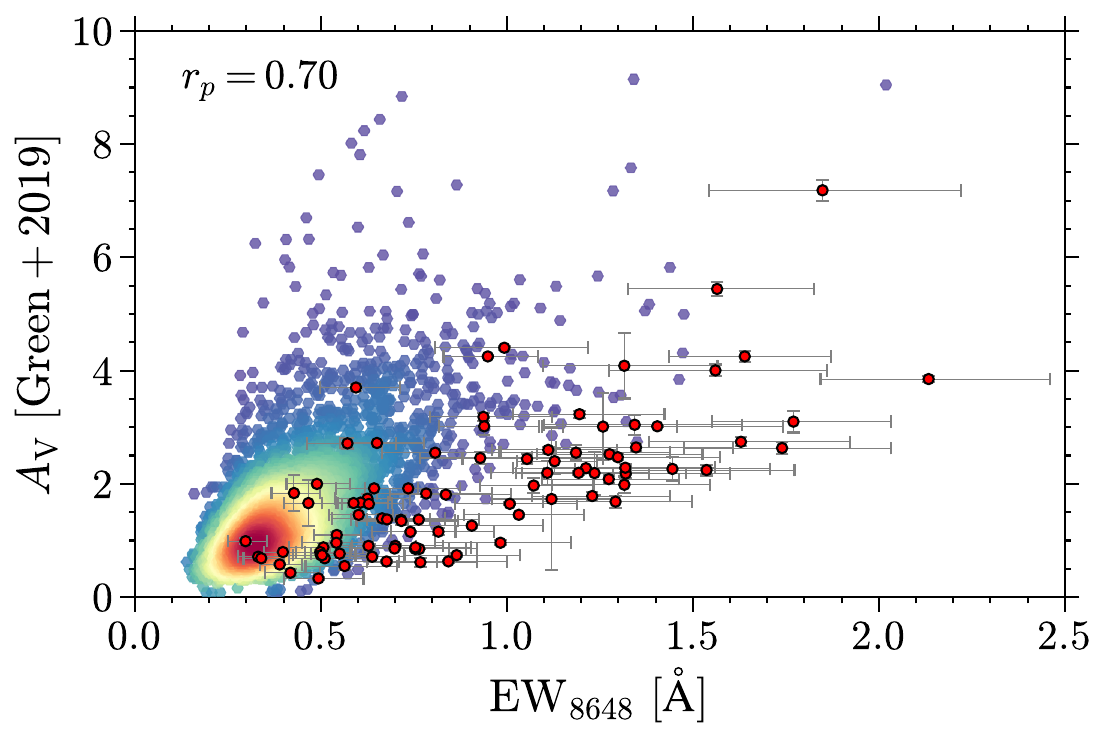}
\caption{Correlation between $\rm EW_{8648}$ and $\Av$ from \citet{Green2019} for 93 selected measurements. The underlying points 
are the results from \citetalias{Schultheis2023FPR}, colored by the number density. The Pearson coefficient ($r_p$) for the red dots 
is indicated as well.} 
\label{fig:ew8648-Av} 
\end{figure}

The DIB signal around 8648\,{\AA} was first reported and measured by \citet{Sanner1978}, with positive supports from \citet{HL1991},
\citet{JD1994}, \citet{Wallerstein2007}, and \citet{Munari2008}, but it was missed in the DIB survey of \citet{Galazutdinov2000a} 
and \citet{Fan2019}. These inconsistent results could be due to the difficulties in measuring a weak and broad DIB feature.
High-resolution spectra of early-type stars used in these previous studies would introduce uncertainties in the continuum placement 
when measuring such broad DIBs \citep{Sonnentrucker2018}. Further, stellar lines would cause contamination even for early-type 
stars, such as the very strong Paschen 13 line (see Fig. 1 in \citealt{Munari2008} for examples) and the \ion{He}{i} line at 
8648.3\,{\AA} reported by \citet{Krelowski2019b} in a B-type star (HD 169454). Under these effects, any conclusions about this DIB 
signal could be distorted in case studies with early-type stars.

Compared to early studies, the large number of Gaia RVS spectra allows us to systematically investigate this signal with a much 
larger spatial coverage. Based on the BNM, we successfully modeled the stellar components of late-type stars and detected the DIB 
signal near 8648\,{\AA} in stacked RVS spectra in \citealt{hz2022} and \citetalias{Schultheis2023FPR}. We cite this DIB as 
$\lambda$8648 following the suggestion in \citet{JD1994}, but we got a smaller $\lambda_0$ as $8645.3\,{\pm}\,1.4$\,{\AA} in 
\citet{hz2022}. The profile of $\lambda$8648 was found to be very shallow and broad. In this work, we further verified that 
DIB\,$\lambda$8648 can be detected in individual RVS spectra although their S/N are much lower than those after stacking. Figure 
\ref{fig:fit-example} already shows the clear profile of $\lambda$8648 in four ISM spectra. In this section, we selected 179 
measurements from the DIB catalog for a statistical study of $\lambda$8648 and its correlation with $\lambda$8621, with strict 
criteria: $\mathcal{D}_{8648}\,{>}\,3R_C$, $8640\,{<}\,\lambda_{8648}\,{<}\,8655$\,{\AA}, $8\,{<}\,\sigma_{8648}\,{<}\,12$\,{\AA}, 
and $\rm S/N\,{>}\,50$. 

The high consistency between fitted and integrated $\rm EW_{8648}$ (Fig. \ref{fig:dib8648} (a)) demonstrates that the DIB profile
of these cases was properly fitted. The general decrease of S/N for large $\rm EW_{8648}$ caused the slight deviation. It should 
be noted that the fitted and integrated $\rm EW_{8648}$ shown in Fig. \ref{fig:dib8648} (a) were simply calculated outside the 
masked region between 8660 and 8668\,{\AA} where the \ion{Ca}{ii} residuals would exist (see Fig. \ref{fig:fit-example} for example). 
So they are smaller than the $\rm EW_{8648}$ shown in Fig. \ref{fig:dib8648} (b) that were directly calculated by the fitted DIB 
parameters. 

$\rm EW_{8621}$ and $\rm EW_{8648}$ present a linear correlation with a Pearson coefficient ($r_p$) of 0.82 (Fig. \ref{fig:dib8648} 
(b)). However, the $\rm EW_{8621}/EW_{8648}$ ratio in this work is systematically smaller than that in \citetalias{Schultheis2023FPR}, 
especially for small $\rm EW_{8648}$. The cause could be a detection bias due to the limited S/N of individual RVS spectra. For
$\mathcal{D}_{8648}$ is only about one third of $\mathcal{D}_{8621}$, weak $\lambda$8648 is harder to be detected at a given 
magnitude of S/N than weak $\lambda$8621, resulting in a lack of measurements with large $\rm EW_{8621}/EW_{8648}$. This effect
should be lighter for \citetalias{Schultheis2023FPR} due to their much higher S/N of the stacked ISM spectra. On the other hand,
after visually checking the extreme measurements with $\rm EW_{8648}\,{\gtrsim}\,0.8$\,{\AA} but $\rm EW_{8621}\,{\lesssim}\,0.2$\,{\AA},
we found that the jagged noise within the very broad profile of $\lambda$8648 could lead to an overestimation of $\rm EW_{8648}$.
There is another possibility. If the carriers of $\lambda$8621 and $\lambda$8648 have different spatial distributions and
$\lambda$8648 carrier is more compact, the stacking of ISM spectra in a 3D volume will obtain a smaller $\rm EW_{8648}$ (lower
mean abundance) compared to $\rm EW_{8621}$. Nevertheless, without a map of their distribution, we cannot analyze the extent of 
this effect. Additionally, $\rm EW_{8621}/EW_{8648}$ would also vary from one sightline to another and present different relations 
with different samples.

The measured central wavelength of the two DIBs also presents a moderate linear relation with $r_p\,{=}\,0.68$ (Fig. \ref{fig:dib8648}
(c)), but their FWHM is noise-dominated, especially for $\lambda$8648. With a linear fit to $\lambda_{8621}$ and $\lambda_{8648}$,
we made a rough estimation of $\lambda_0$ for $\lambda$8648 as 8646.31\,{\AA} in vacuum, assuming that the carriers of 
$\lambda$8621 and $\lambda$8648 are comoving. This value corresponds to 8643.93\,{\AA} in air, which is much smaller than
previous suggestions, such as 8650\,{\AA} by \citet{Sanner1978}, 8648.28\,{\AA} by \citet{JD1994}, and 8649\,{\AA} by \citet{HL1991}.
The difference between this result and \citet{hz2022} is 1.01\,{\AA}, slightly larger than the mean uncertainty of $\lambda_{8648}$ 
of the used cases as 0.69\,{\AA}. Figure \ref{fig:ew8648-Av} shows the correlation between $\rm EW_{8648}$ and $\Av$ from 
\citet{Green2019} for 93 cases (others are out of the sky coverage of \citealt{Green2019}). A moderate linear correlation can be
found with $r_p\,{=}\,0.70$. Compared to $\lambda$8621, DIB\,$\lambda8648$ presents a worse correlation with dust reddening, which 
has been noted in \citet{hz2022} and \citetalias{Schultheis2023FPR}. Similar to $\rm EW_{8621}/EW_{8648}$, the $\Av/{\rm EW_{8648}}$ 
ratio in this work is systematically smaller than that in \citetalias{Schultheis2023FPR}.

\subsection{Reassess the DIB measurements in Gaia DR3} \label{subsect:GDR3}

\begin{figure*}
\centering
\includegraphics[width=16.8cm]{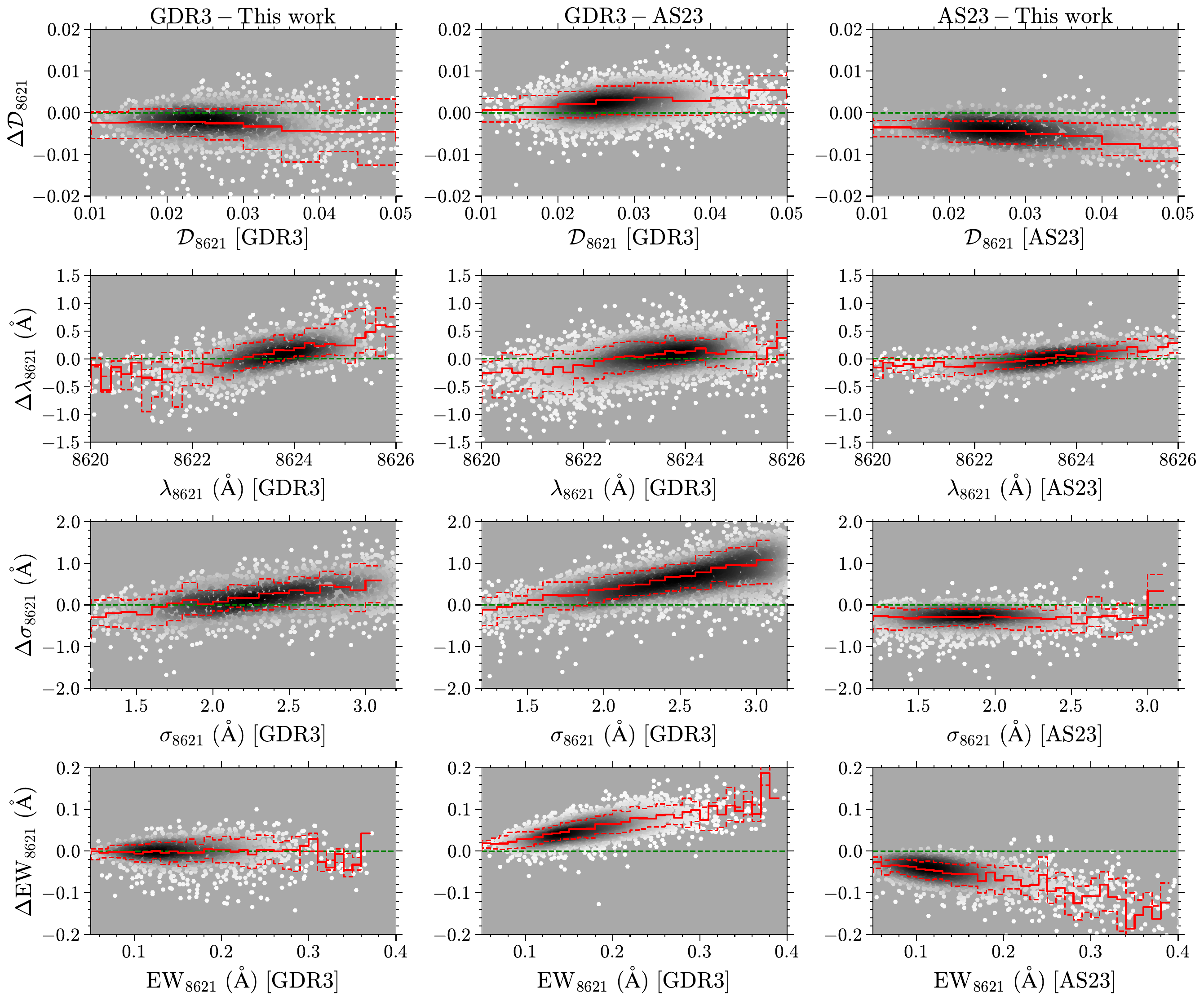}
\caption{Difference in DIB parameters ($\mathcal{D}_{8621}$, $\lambda_{8621}$, $\sigma_{8621}$, and $\rm EW_{8621}$) between 
\citetalias{Schultheis2023DR3}, \citetalias{Saydjari2023}, and this work as a function of the fitted values for the joint samples. 
The gray scale indicates the number density of the data points estimated by the Gaussian KDE. The data are binned with a step of 
0.005 for $\mathcal{D}_{8621}$, 0.2\,{\AA} for $\lambda_{8621}$, 0.1\,{\AA} for $\sigma_{8621}$, and 0.01\,{\AA} for $\rm EW_{8621}$, 
respectively. The solid red lines in each panel represent the median differences in each bin and the two dashed red lines show the 
16th and 84th percentiles. }
\label{fig:GDR3} 
\end{figure*}

Considering the small distances to the background stars (median distance is 1.31\,kpc) and the moderate S/N of the RVS spectra 
(median S/N is 115.3) for the HQ sample of \citetalias{Schultheis2023DR3}, its measured $\lambda_{8621}$ and $\sigma_{8621}$ should
present a quasi-Gaussian distribution centered on $\lambda_0$ and the mean Gaussian width with a dispersion due to the uncertainties
and the Galactic rotation (like the distribution seen in Fig. \ref{fig:lambda-sigma} for this work and \citetalias{Saydjari2023}).
However, a strong dependence between $\lambda_{8621}$ and $\sigma_{8621}$ is clearly seen for the HQ sample of \citetalias{Schultheis2023DR3}
(see Fig. 1 in \citetalias{Saydjari2023}) which would be attributed to the improperly modeled stellar lines. On one hand, for small 
$\sigma_{8621}$ ($\lesssim$1\,{\AA}) and large $\lambda_{8621}$ (around 8624--8626\,{\AA}), the fittings could be purely pseudo 
for the residuals of \ion{Fe}{i} lines there would be stronger than the DIB features. On the other hand, the increase of $\sigma_{8621}$
with $\lambda_{8621}$ shifting shortward (8622--8624\,{\AA}) implies a broadening of the DIB profile caused by the noise and 
stellar residuals as more stellar lines at shorter wavelength in the vicinity of the DIB signal. 

Compared to \citetalias{Schultheis2023DR3}, the ISM spectra derived by the data-driven methods in this work and \citetalias{Saydjari2023}
are less influenced by the stellar residuals. Therefore, to estimate the magnitude of the biases caused by the stellar residuals, 
we compare the DIB parameters from \citetalias{Schultheis2023DR3} to those from this work and \citetalias{Saydjari2023} fitted in 
the same RVS spectra, specifically 1518 cases between \citetalias{Schultheis2023DR3} and this work and 3167 cases between 
\citetalias{Schultheis2023DR3} and \citetalias{Saydjari2023}, as well as 2000 cases between this work and \citetalias{Saydjari2023} 
as a control group. We only consider the highest level of the DIB quality flag in \citetalias{Schultheis2023DR3} (i.e. 
$\rm QF\,{=}\,0$, see Sect. 2 in \citetalias{Schultheis2023DR3} and Sect. 8.9 in \citealt{Recio-Blanco2023} for details). The 
differences in DIB parameters as a function of the fitted values are shown in Fig. \ref{fig:GDR3}, and their statistics, including 
the median difference (MED), the root-mean-square difference (RMSD), and the absolute difference not exceeded by 90\% of sources 
(AD90), are presented in Table \ref{tab:statistics}. The $\mathcal{D}_{8621}$ for \citetalias{Saydjari2023} (not given in their
catalog) was calculated by their $\lambda_{8621}$ and $\rm EW_{8621}$ with a Gaussian function. 

The impact of the stellar residuals causes a systematic shift of $\lambda_{8621}$ in \citetalias{Schultheis2023DR3}, which is
clearly seen as the systematic variation of $\Delta \lambda_{8621}$ with $\lambda_{8621}$ in Fig. \ref{fig:GDR3}. This phenomenon
coincides with the $\lambda_{8621}-\sigma_{8621}$ dependence discussed above. The MED of $\Delta \lambda_{8621}$ is much larger 
for this work (0.073\,{\AA}) than for \citetalias{Saydjari2023} (0.018\,{\AA}), but the RMSD is similar (0.35\,{\AA}) and close 
to the pixel size, corresponding to $12\,\kms$. This value is comparable to the mean uncertainty of $\lambda_{8621}$ (0.376\,{\AA}) 
in \citetalias{Schultheis2023DR3} for the joint samples as well. Considering AD90, the maximum shift for most $\lambda_{8621}$ is
about 0.56\,{\AA} ($\sim$19\,$\kms$), 1.5 times larger than the mean uncertainty. As a comparison, $\lambda_{8621}$ in this work
and in \citetalias{Saydjari2023} are highly consistent with each other with a median difference of only 0.008\,{\AA} and with a 
halved RMSD and AD90. Nevertheless, $\Delta \lambda_{8621}$ between \citetalias{Saydjari2023} and this work also presents a weak 
dependence on $\lambda_{8621}$, which could be due to a weak stellar impact, despite most $\Delta \lambda_{8621}$ are smaller than 
the RVS wavelength pixel size used in this work. 

\begin{table}
\centering
\caption{Statistics of the differences in DIB parameters between \citetalias{Schultheis2023DR3}, \citetalias{Saydjari2023}, 
and this work.}
\label{tab:statistics}
\begin{tabular}{lrcc}
\hline\hline 
       & MED & RMSD & AD90  \\ [0.05ex]
\hline   
GDR3 -- This work:   & & & \\ 
$\mathcal{D}_{8621}$     & --0.002 & 0.006 & 0.009 \\
$\lambda_{8621}$ ({\AA}) & 0.073   & 0.353 & 0.564 \\
$\sigma_{8621}$ ({\AA})  & 0.164   & 0.469 & 0.773 \\
$\rm EW_{8621}$ ({\AA})  & --0.002 & 0.030 & 0.046 \\ [0.05ex]
\hline
GDR3 -- AS23:   & & & \\  
$\mathcal{D}_{8621}$     & 0.002   & 0.004 & 0.007 \\
$\lambda_{8621}$ ({\AA}) & 0.018   & 0.342 & 0.546 \\
$\sigma_{8621}$ ({\AA})  & 0.557   & 0.752 & 1.221 \\
$\rm EW_{8621}$ ({\AA})  & 0.050   & 0.061 & 0.093 \\ [0.05ex]
\hline
AS23 -- This work:   & & & \\ 
$\mathcal{D}_{8621}$     & --0.005 & 0.010 & 0.012 \\
$\lambda_{8621}$ ({\AA}) & 0.008   & 0.183 & 0.292 \\
$\sigma_{8621}$ ({\AA})  & --0.293 & 0.433 & 0.663 \\
$\rm EW_{8621}$ ({\AA})  & --0.050 & 0.094 & 0.118 \\ [0.05ex]
\hline
\end{tabular}
\tablefoot{Columns: the median difference (MED), the root-mean-square difference (RMSD), and the absolute difference not exceeded 
by 90\% of sources (AD90).}
\end{table}

The overestimated $\sigma_{8621}$ in \citetalias{Schultheis2023DR3} has a MED of 0.164\,{\AA} compared to this work and tends
to become larger with the increase of $\sigma_{8621}$. The RMSD (0.469\,{\AA}) is slightly larger than the mean uncertainty of
$\sigma_{8621}$ in \citetalias{Schultheis2023DR3} (0.405\,{\AA}) and the AD90 reaches over 0.7\,{\AA}. As a comparison, $\sigma_{8621}$ 
measured in this work is larger than that in \citetalias{Saydjari2023} with a nearly constant difference (a MED of ${-}0.293$\,{\AA}). 
Besides the overestimated $\sigma_{8621}$, $\mathcal{D}_{8621}$ in \citetalias{Schultheis2023DR3} is oppositely smaller than that 
in this work, and $\Delta \mathcal{D}_{8621}$ presents an increasing trend with $\mathcal{D}_{8621}$ as well. Additionally, $\rm 
EW_{8621}$ are highly consistent with each other between \citetalias{Schultheis2023DR3} and this work, with a MED of ${-}0.002$\,{\AA} 
and a RMSD of 0.030\,{\AA}, comparable to the mean EW uncertainty (0.024\,{\AA}) in \citetalias{Schultheis2023DR3}. Overall, we 
propose that the impact of stellar residuals led to a distortion of the DIB profile in \citetalias{Schultheis2023DR3}, slightly
becoming shallower and broadening. The center of the profile was also shifted to one or two pixels at most for the joint sample, 
but the area of the profile ($\rm EW_{8621}$) remained unchanged. 

The $\rm EW_{8621}$ in this work is systematically larger than that in \citetalias{Saydjari2023}, and the difference further 
increases with the fitted values and can reach around 0.1\,{\AA} for $\rm EW_{8621}\,{\sim}\,0.3$\,{\AA}. The mean $\Delta \rm 
EW_{8621}$ is 27.1\% relative to our measurements, much larger than the mean uncertainty of $\rm EW_{8621}$ (12.9\%). Since 
\citetalias{Saydjari2023} and this work have consistent $\lambda_{8621}$ and nearly constant $\Delta \sigma_{8621}$, the rising 
of $\Delta \rm EW_{8621}$ would be caused by different ML algorithms that model the DIB depth in different ways. Moreover, 
\citetalias{Saydjari2023} modeled the profile of $\lambda$8621 with a Gaussian function, but we added a Lorentzian function for 
$\lambda$8648 and a linear continuum accounting for RVS spectra with ill normalization. Nevertheless, we note that \citetalias{
Schultheis2023DR3} made use of synthetic spectra from stellar models and a simple Gaussian fitting, but their $\rm EW_{8621}$ is 
highly consistent with that in this work. Thus, the influence of the DIB model should be not so significant. The last factor is 
the fitting technique. Specifically, the ISM spectrum was first derived in this work, and then the DIB profile was modeled. While 
in \citetalias{Saydjari2023}, the DIB profile was implemented as a pixel-by-pixel covariance matrix, together with the stellar 
components and the noise, and was optimized in a set of grids. Despite the systematic difference in $\rm EW_{8621}$, the span of 
the 16th to the 84th percentiles of $\Delta \rm EW_{8621}$ (a measure of the magnitude of the dispersion deducting the tendency) 
is similar to that between \citetalias{Schultheis2023DR3} and this work and that between \citetalias{Saydjari2023} and this work 
for $\rm EW_{8621}\,{\lesssim}\,0.3$\,{\AA}. 

\section{Discussion} \label{sect:discuss}

\subsection{DIB\texorpdfstring{$\,\lambda$}{l}8621 correlates better with neutral than molecular hydrogen} \label{subsect:dib-CO-HI}

Motivated by the good consistency and the multimodality found between DIB\,$\lambda$8621 and $^{12}$CO in velocity structures, 
\citetalias{Saydjari2023} directly compared $\Vdib$ and $V_{\rm CO}$ by a peak finding method. Specifically, signals of $\lambda$8621
and $^{12}$CO were simply matched by the position of the background stars and the space grid of $^{12}$CO map (a resolution of 
$0.125^{\circ}$ for \citealt{Dame2001}). Then for any detected $^{12}$CO emission within 1$\sigma$ of $\Vdib$, $\Vdib$ was compared
to the intensity-weighted $V_{\rm CO}$ calculated within nine velocity channels around $\Vdib$ (see Sect. 3.4.2 and Appendix E in
\citetalias{Saydjari2023} for details). With a linear fit restricted to $|V_{\rm CO}|\,{<}\,25\,\kms$, they got a slope of 0.95
suggesting that the DIB\,$\lambda$8621 carrier and $^{12}$CO are comoving and a small intercept of ${-}0.52\,\kms$ validating their
$\lambda_0$ estimation.

\begin{figure}
\centering
\includegraphics[width=8cm]{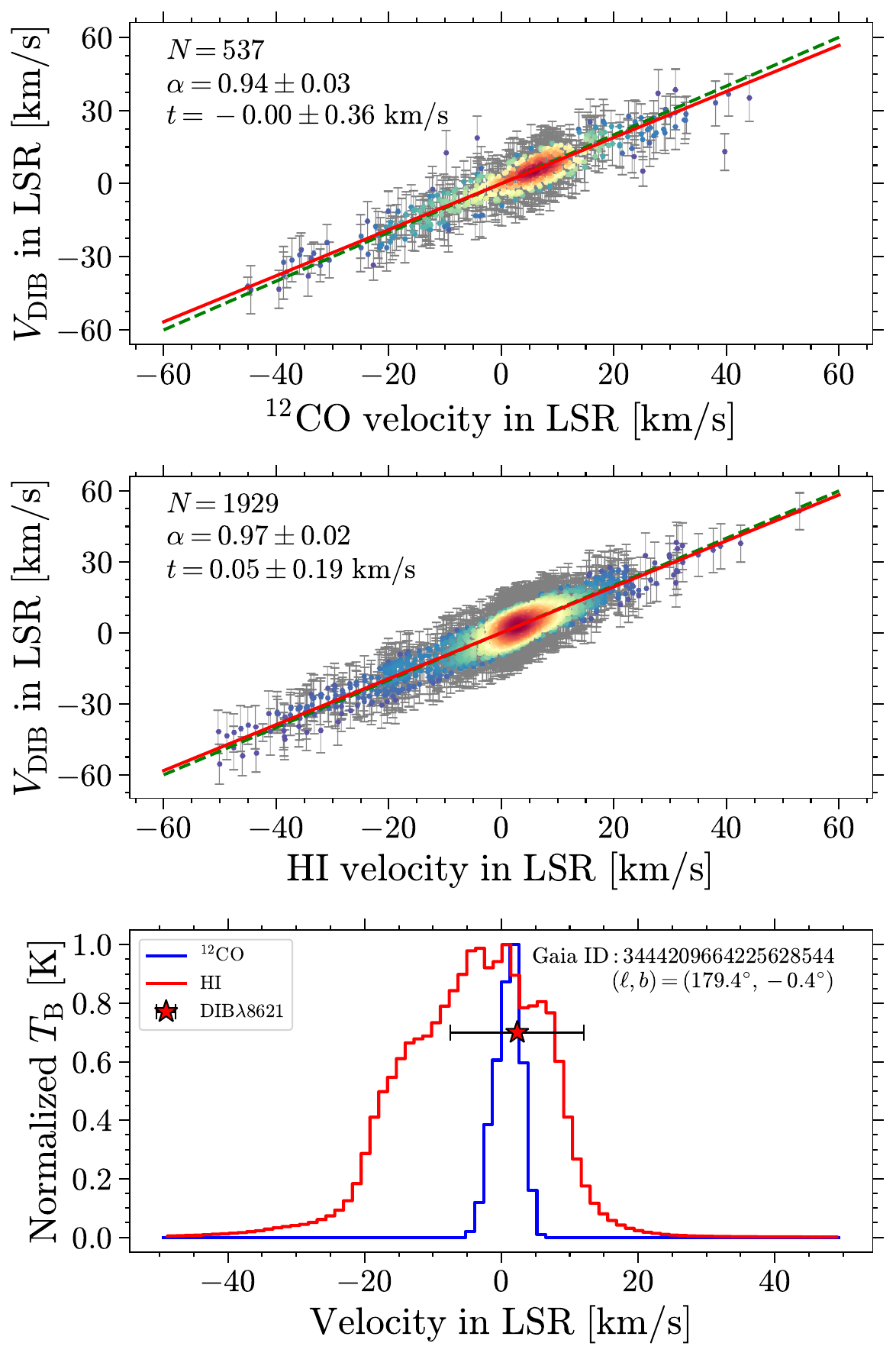}
\caption{{\it Upper and middle panels:} Correlation between the DIB\,$\lambda$8621 velocity ($\Vdib$) and the intensity-weighted 
velocity of the nearest $^{12}$CO \citep{Dame2001} and \ion{H}{i} \citep{HI4PI2016} component, respectively. The color scale indicates 
the number density of the data points estimated by the Gaussian KDE. The error of $\Vdib$ is simply estimated by the uncertainty 
of $\lambda_{8621}$. The red line is the linear fit to the LSR velocities. The fitted slope ($\alpha$) and intercept ($t$), as well 
as the number of the points, are indicated. The dashed green line traces the one-to-one correspondence. {\it Lower panel:} An example 
to illustrate the peak finding method (see Sect. \ref{subsect:dib-CO-HI}). The red and blue lines are the spectra of $^{12}$CO and 
\ion{H}{i}, respectively, toward the same sightline. The red star indicates the LSR velocity of the DIB signal detected in this sightline.
The Galactic coordinate ($\ell,b$) and the Gaia source ID of the background star are marked.} 
\label{fig:DCH1} 
\end{figure}

We fellow their peak finding method but only considered the $V_{\rm CO}$ at the peak temperature of $^{12}$CO toward each sightline 
and made use of 7267 DIB measurements with $|b|\,{<}\,30^{\circ}$, ${\rm err}(\lambda_{8621})<1$\,{\AA}, and ${\rm err}(\Vstar)<5\,\kms$.
We further compared $\Vdib$ to $V_\ion{H}{i}$ using \ion{H}{i} data from HI4PI \citep{HI4PI2016}. Finally, we found 537 cases with
matched $\lambda$8621 and $^{12}$CO in velocity and 1929 for $\lambda$8621 and \ion{H}{i}. With a cross-match in velocity, it is not 
surprising to find a strong one-to-one relationship between $\lambda$8621 and $^{12}$CO as in \citetalias{Saydjari2023}, as well as 
between $\lambda$8621 and \ion{H}{i} even with a slope closer to 1 (see the upper and middle panels in Fig. \ref{fig:DCH1}). The lower 
panel in Fig. \ref{fig:DCH1} shows an example to illustrate the peak finding method. It can be found that the $^{12}$CO emission is 
narrow and compact mainly within ${\pm}5\,\kms$, while the uncertainty of $\Vdib$ (9.78\,$\kms$) is much larger than the velocity 
span of $^{12}$CO. On the other hand, the \ion{H}{i} emission covers a much wider velocity range and contains multiple components 
that cannot be resolved in the RVS spectra. With the limited accuracy of $\Vdib$ and the strong bias of the peak finding method, 
it is hard to conclude that the perfect association between $^{12}$CO and the carrier of DIB\,$\lambda$8621 implies a clumpiness 
of the DIB carrier. The associated velocity between DIB\,$\lambda$8621 and $^{12}$CO, as well as \ion{H}{i}, is more like a result 
of the general Galactic rotation of these gaseous ISM species at a similar distance.

\begin{figure}
\centering
\includegraphics[width=8cm]{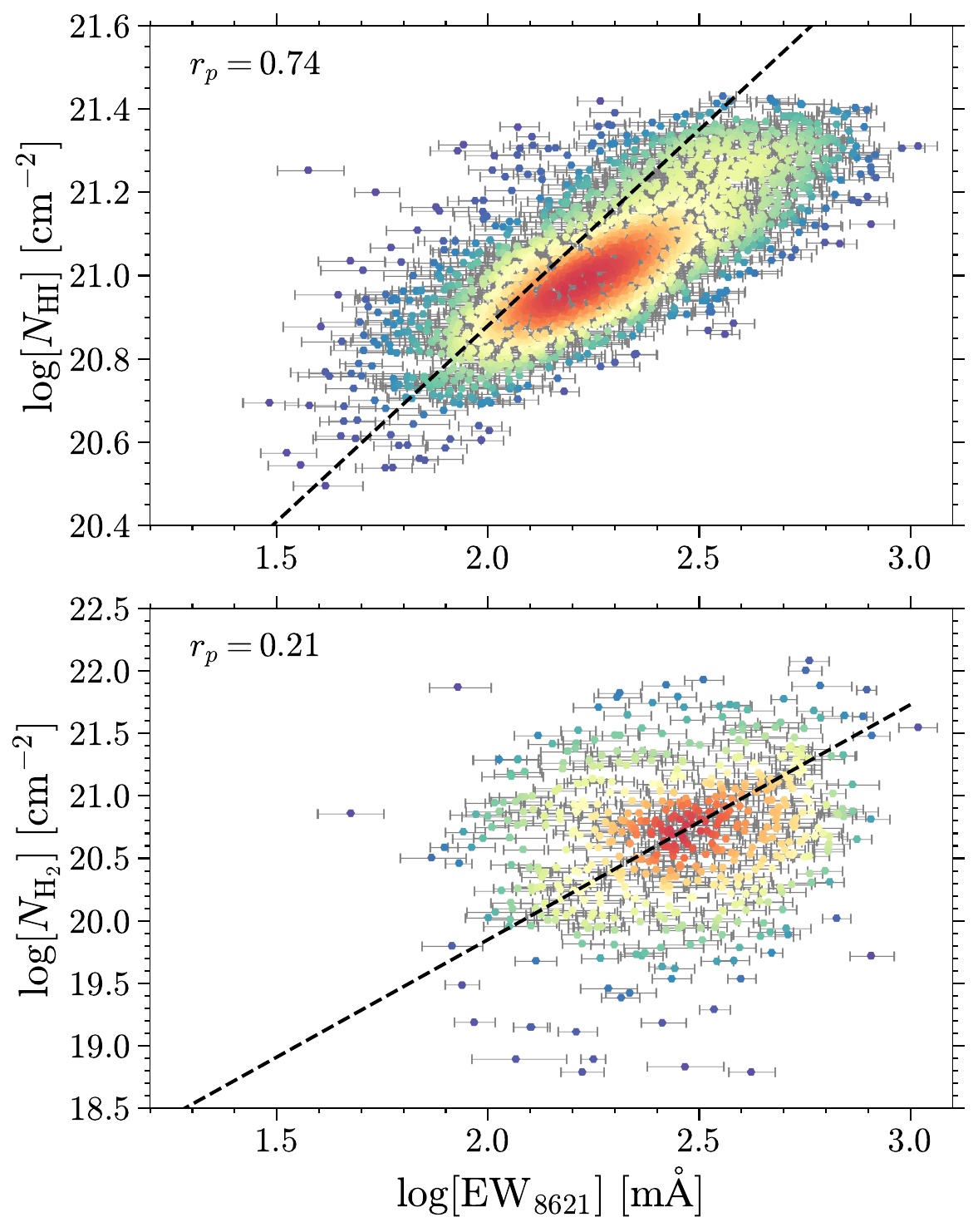}
\caption{Correlation between $\rm EW_{8621}$ and $\NHI$ ({\it upper panel}) and between $\rm EW_{8621}$ and $\NHt$ ({\it lower panel}) 
in a logarithmic scale. The Pearson coefficient ($r_p$) is indicated. The color scale indicates the number density. The dashed
lines are the linear fit results from \citet{Friedman2011} but for DIB\,$\lambda$5780. } 
\label{fig:DCH2} 
\end{figure}

Based on the velocity-matched $\rm DIB\,{-}\,^{12}$CO and DIB\,--\,\ion{H}{i} pairs, we made a coarse investigation of the correlation 
between the DIB strength and the hydrogen abundance for both neutral hydrogen ($\NHI$) and molecular hydrogen ($\NHt$). We calculated 
their abundance as $\NHI = 1.823 \times 10^{18} \times I_\ion{H}{i}$ \citep{HI4PI2016} and $\NHt = 2 \times 10^{20} \times W_{\rm CO}$ 
\citep{Bolatto2013}, where $I_\ion{H}{i}$ and $W_{\rm CO}$ are the velocity-integrated intensity calculated with nine velocity 
channels around their matched $\Vdib$. This analysis was based on an assumption that ISM species with similar radial velocities 
are mainly located at a similar distance, so that the DIB features can be compared to the corresponding \ion{H}{i} and $^{12}$CO 
emission, with a narrow-range integration to deduct the foreground and background contamination. This is certainly an ideal 
assumption. As shown in Fig. \ref{fig:DCH2} in a logarithmic scale, a moderate linear correlation has been found between 
$\rm EW_{8621}$ and $\NHI$ ($r_p\,{=}\,0.74$), while $\rm EW_{8621}$ is not sensitive to $\NHt$ ($r_p\,{=}\,0.21$). Therefore, 
the carrier of $\lambda$8621 would correlate much better with neutral hydrogen than molecular hydrogen. Although $\rm EW_{8621}$ 
is proportional to the column density of the carrier only between the background star and us and the \ion{H}{i} and $^{12}$CO 
observations may trace the hydrogen abundance in a much wider distance range, the narrow integration range around $\Vdib$ seems 
to alleviate this influence.

A set of strong optical DIBs have been reported to tightly correlate with $\NHI$ but only present a loose correlation with $\NHt$
when $\NHt\,{>}\,10^{20}\,{\rm cm^{-2}}$ \citep[e.g.][]{Herbig1993,Friedman2011,Lan2015}. The relationship between $\rm EW_{8621}$ 
and $\NHI$ revealed by our coarse analysis corresponds to this inference. Particularly, \citet{Friedman2011} derived a tight
correlation between DIB\,$\lambda$5780 and $\NHI$, and the range of $\NHI$ where they found the correlation ($20\,{\sim}\,21.5\,{\rm 
cm^{-2}}$) is similar to ours (see the dashed black line in the upper panel of Fig. \ref{fig:DCH2}). According to \citet{Fan2019},
the $\EBV$-normalized EW of $\lambda$5780 is twice as much as that of $\lambda$8621. Hence, the larger $\rm EW_{8621}$ than $\rm 
EW_{5780}$ at a given $\NHI$ seen in Fig. \ref{fig:DCH2} would be caused by the underestimation of $\NHI$ in our analysis due to 
the narrow integration range. Nevertheless, we did not find a loose correlation between $\rm EW_{8621}$ and $\NHt$ even for 
$\NHt\,{>}\,10^{20}\,{\rm cm^{-2}}$, although the fitted line of the ${\rm EW_{5780}}\,{-}\,\NHt$ relation in \citet{Friedman2011} 
crosses with the highest density region of our sample for ${\rm log[EW_{8621}]}$ and ${\rm log}[\NHt]$ (see the dashed black line 
in the lower panel of Fig. \ref{fig:DCH2}). The possible variation of the $X_{\rm CO}$ factor and the saturation problem of $^{12}$CO 
would further hamper the investigation of the correlation between $\rm EW_{8621}$ and $\NHt$.

\begin{figure}
\centering
\includegraphics[width=8cm]{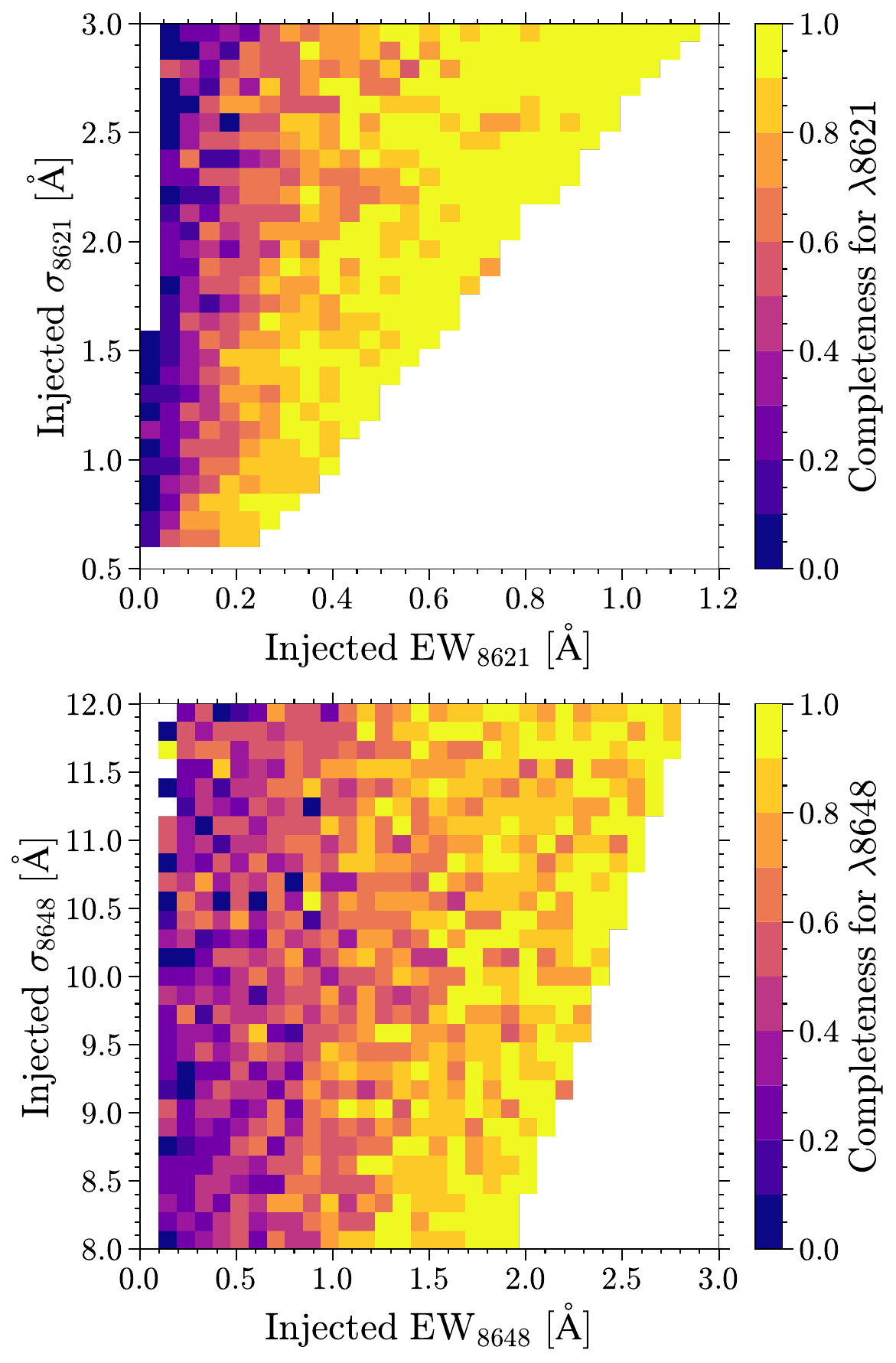}
\caption{Distribution of the estimated completeness for DIBs $\lambda$8621 ({\it upper panel}) and $\lambda$8648 ({\it lower panel}), 
respectively, as a function of EW and $\sigma_{\rm DIB}$ based on the results of the injection tests (see Appendix \ref{appsect:intest}).} 
\label{fig:completeness} 
\end{figure}

\subsection{Completeness of the DIB catalog} \label{subsect:completeness}

With the injection test, \citetalias{Saydjari2023} estimated the completeness of their DIB catalog by selecting the ``good'' 
measurements having $\Delta \mathcal{D}\,{<}\,20\%$, $\Delta \sigma_{\rm DIB}\,{<}\,20\%$, and $\Delta \lambda_{\rm DIB}\,{<}\,2.5$\,{\AA} 
(differences between fitted and injected values, see Appendix F in \citetalias{Saydjari2023}). For our selected DIB catalog, the mean 
uncertainty of $\mathcal{D}_{8621}$ and $\sigma_{8621}$ is about 10\% and the $\Delta \lambda_{8621}$ is mainly within 0.5 between 
this work and \citetalias{Schultheis2023DR3} (see Table \ref{tab:statistics} and Sect. \ref{subsect:GDR3}). Hence, we estimated the 
completeness of our DIB catalog with more rigorous criteria, that is $\Delta \mathcal{D}\,{<}\,10\%$, $\Delta \sigma_{\rm DIB}\,{<}\,10\%$, 
and $\Delta \lambda_{\rm DIB}\,{<}\,0.5$\,{\AA}, based on the results of the injection test (see Appendix \ref{appsect:intest}). The 
distribution of the estimated completeness for DIB\,$\lambda$8621 as a function of the injected $\rm EW_{8621}$ and $\sigma_{8621}$ 
(upper panel in Fig. \ref{fig:completeness}) is similar to that in \citetalias{Saydjari2023}. The completeness generally decreases 
with $\rm EW_{8621}$, but its variation with $\sigma_{8621}$ is not so clear as in \citetalias{Saydjari2023} due to our smaller sample
used in the injection test. In our DIB catalog, the median $\rm EW_{8621}$ is 0.2\,{\AA} and its 16th and 84th percentiles are 0.1 
and 0.4\,{\AA}. The mean completeness at $\rm EW_{8621}\,{\sim}\,0.2$\,{\AA} is about 65\% and between 0.1 and 0.4\,{\AA} is 68\%. 
This is an optimistic estimate as the injection test was simple and ideal. Nevertheless, this percentile is still much larger than 
the fraction of the joint sample to the total between this work and \citetalias{Saydjari2023} ($\sim$25\%), which could be a result 
of the selection bias on the DIB catalog (see Appendix \ref{appsect:select-bias} for a detailed discussion).

The estimated completeness for DIB\,$\lambda$8648 has a much weaker dependence on $\rm EW_{8648}$ and $\sigma_{8648}$, indicating 
a stronger influence of the correlated noise and stellar residuals on the completeness of DIB\,$\lambda$8648 measurements. At $\rm 
EW_{8648}\,{\sim}\,0.5$\,{\AA}, the mean completeness is only 25\%. If we used loose criteria, the same as those in \citetalias{
Saydjari2023}, the completeness will increase to 53\%. Although the total number of the spectra in the target sample that truly 
contain DIB\,$\lambda$8648 signals is unknown, apparently, we did not get ``enough'' DIB\,$\lambda$8648 measurements (only 179 
after quality filtering) even in the selected DIB catalog. One possible reason is that the selection criteria are too rigorous for 
DIB\,$\lambda$8648. Another is that DIB\,$\lambda$8648 does not exist in every spectrum with DIB\,$\lambda$8621 signals, that is 
the detectable DIB\,$\lambda$8648 sightlines are less than that for DIB\,$\lambda$8621. The completeness for DIB\,$\lambda$8648 
would be overestimated as well.

Based on the estimated completeness and the comparison to \citetalias{Saydjari2023} (Appendix \ref{appsect:select-bias}), the DIB 
catalog in this work would have high purity but low completeness. It is a critical problem to increase both the purity and the 
completeness of the DIB catalog. However, the present RF model cannot make a meaningful estimation of the uncertainty of its 
predictions, and this is a common failure for the ML approach. We will seek more models and more intelligent injection tests or 
simulations in the following works.

\section{Summary and Conclusions} \label{sect:conclusion}

In this work, we developed a Random Forest model to build the stellar templates within the DIB window (8600--8680\,{\AA}) 
using the part of the spectra outside the window. This method can be treated as an improved best-neighbor method developed in 
\citet{Kos2013} and was applied to the RVS spectra published in Gaia DR3. The training set was constituted by 21\,974 spectra 
with $\EBV\,{\leqslant}\,0.02$\,mag, $|b|\,{\geqslant}\,30^{\circ}$, and $\rm S/N\,{>}\,50$. After subtracting the stellar 
components by the generated templates, we fitted DIB\,$\lambda$8621 by a Gaussian function and DIB\,$\lambda$8648 by a Lorentzian 
function, as well as a linear continuum, for 780 thousand target spectra. These target spectra have a mean S/N of 58 and 90\% of 
the spectral S/N is below 116. The mean distance of the background stars is 1.58\,kpc, and 90\% of them are located within 3.61\,kpc.

Considering $\chi^2_{\rm dof}$, noise level (S/N and $R_C$), and the constraints on $\lambda_{8621}$ and $\sigma_{8621}$, we selected
7619 reliable measurements for DIB\,$\lambda$8621 (the DIB catalog can be accessed via the CDS database). Their $\rm EW_{8621}$ 
presented a moderate linear correlation with dust reddening from both \citet{Andrae2023} and \citet{Green2019}, and the mean ${\rm 
EW_{8621}}/\Av$ ratio was consistent with our previous results in \citetalias{Schultheis2023DR3} and \citetalias{Schultheis2023FPR}. 
Using 77 DIB measurements toward the GAC and an assumption of a circular orbit, we determined an updated rest-frame wavelength of 
DIB\,$\lambda$8621 as $\lambda_0\,{=}\,8623.141\,{\pm}\,0.030$\,{\AA} in vacuum, corresponding to 8620.766\,{\AA} in air, 
which was perfectly consistent with the result in \citetalias{Saydjari2023} but bluer than that in \citetalias{Schultheis2023DR3}. 
Calculated by $\lambda_0$, $\Vdib$ in LSR showed a wave pattern with the Galactic longitude, revealing the projected Galactic 
rotation of the carrier of DIB\,$\lambda$8621. The median $\Vdib$ also correlated with the $^{12}$CO velocity structures in the 
local region, especially for the outer Galactic disk. With the peak finding method used in \citetalias{Saydjari2023} and a narrow-range 
integration, we compared $\rm EW_{8621}$ to the neutral ($\NHI$, from \citealt{HI4PI2016}) and molecular ($\NHt$, represented by 
$^{12}$CO) hydrogen column densities. This was a coarse analysis, but it can be found that $\rm EW_{8621}$ correlated much better 
with $\NHI$ ($r_p\,{=}\,0.74$) than $\NHt$ ($r_p\,{=}\,0.21$), which was consistent with the conclusions for strong optical DIBs 
in previous studies.

With rigorous quality control, we obtained 179 reliable measurements of DIB\,$\lambda$8648 in individual RVS spectra, which further 
confirmed this very broad DIB feature. Its EW and central wavelength both presented a moderate linear relation with those of 
DIB\,$\lambda$8621. The $\lambda_0$ of DIB\,$\lambda$8648 was estimated as 8646.31\,{\AA} in vacuum, corresponding to 
8643.93\,{\AA} in air, assuming that the carriers of $\lambda$8621 and $\lambda$8648 are co-moving.

By comparing the DIB parameters in \citetalias{Schultheis2023DR3}, in \citetalias{Saydjari2023}, and in this work, we confirmed the
impact of the stellar residuals on the DIB measurements in Gaia DR3 argued by \citetalias{Saydjari2023}. The stellar impact leads
to a distortion of the DIB profile, resulting in an underestimation of $\mathcal{D}_{8621}$ and an overestimation of $\sigma_{8621}$.
The center of the DIB profile could be also shifted ($\lesssim$0.5\,{\AA}), but $\rm EW_{8621}$ was consistent with our new measurements 
in this work with a median difference of only ${-}0.002$\,{\AA} and a RMSD of 0.030\,{\AA}. 

$\rm EW_{8621}$ in this work is systematically larger than that in \citetalias{Saydjari2023} and the difference further increases
with the fitted EW. The cause could be the different ML algorithms and fitting techniques used in the two works. The selection bias
of the DIB catalog was clearly revealed by the crossed groups between \citetalias{Saydjari2023} and this work. The DIB catalog has
high purity but low completeness. In the following works, we will apply more ML algorithms to different survey data and investigate
their consistency and/or systematic differences. 

\begin{acknowledgements}
We thank the anonymous referee for very helpful suggestions and constructive comments.
This work has made use of data from the European Space Agency (ESA) mission Gaia (\url{https://www.cosmos.esa.int/gaia}), 
processed by the Gaia Data Processing and Analysis Consortium (DPAC, \url{https://www.cosmos.esa.int/web/gaia/dpac/consortium}). 
Funding for the DPAC has been provided by national institutions, in particular, the institutions participating in the Gaia Multilateral Agreement.
This work is supported by the National Natural Science Foundation of China (grant No. 12203099).
HZ is funded by the China Postdoctoral Science Foundation (No. 2022M723373) and the Jiangsu Funding Program for Excellent Postdoctoral Talent.
HZ acknowledges the helpful discussions with Dr. Jianjun Chen, Dr. Biwei Jiang, and Xiaoxiao Ma. 
TZ acknowledges financial support of the Slovenian Research Agency (research core funding No. P1-0188) and the European Space Agency
(Prodex Experiment Arrangement No. 4000142234).
\end{acknowledgements}

\bibliographystyle{aa}
\bibliography{reference.bib}

\appendix

\section{Performance of the validation set}

\begin{figure*}
  \centering
  \includegraphics[width=16.8cm]{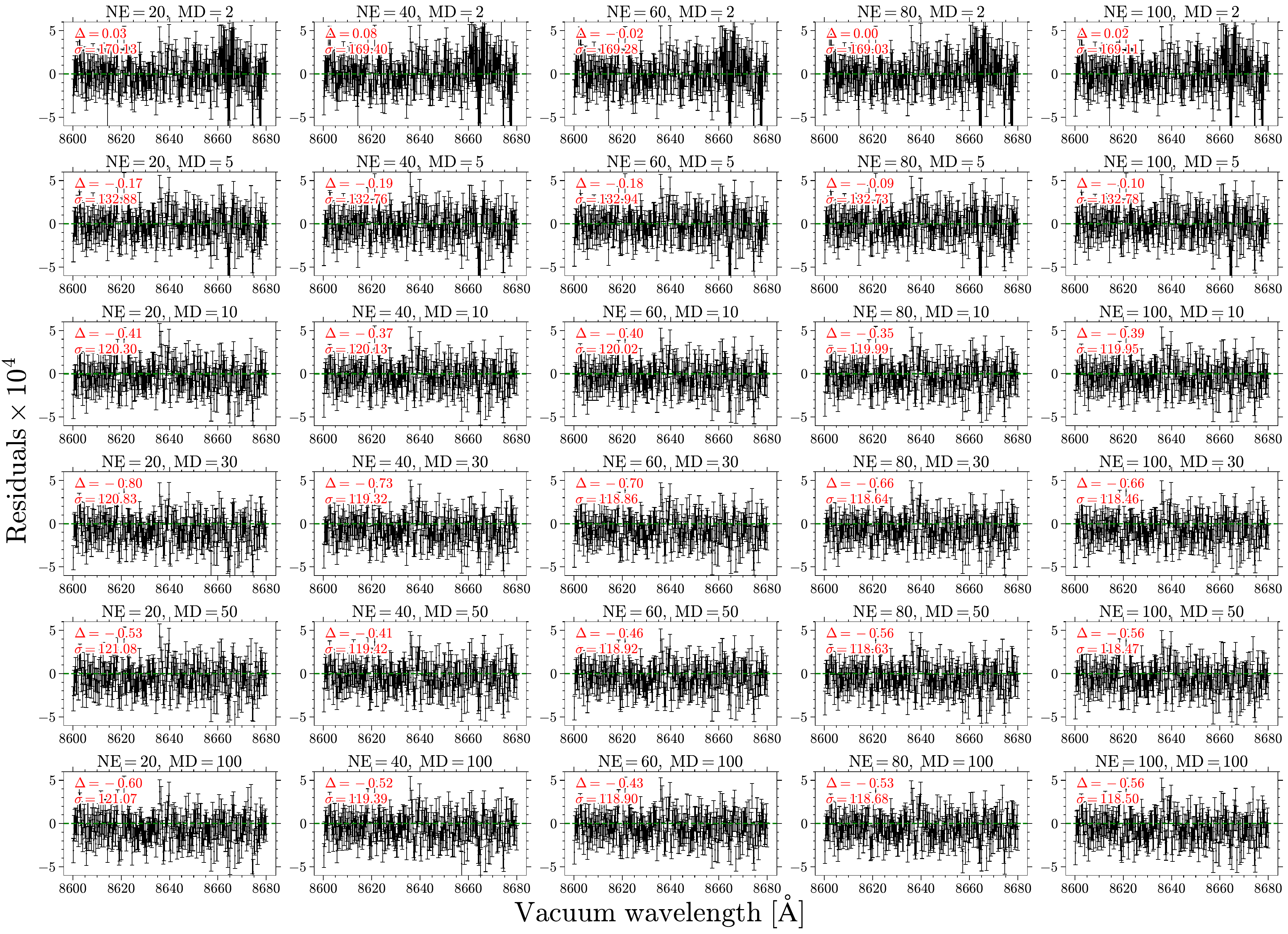}
  \caption{The mean residuals between observed and modeled normalized fluxes taken for 7324 RVS spectra in the validation set
  at each wavelength pixel in the DIB window as a function of the spectral wavelength for different parameter pairs of `n\_estimators' 
  (NE) and `max\_depth' (MD) in the RF model ({\it scikit-learn}). The error bar represents the standard deviation of the residuals 
  of individual spectra at each wavelength pixel. The mean value ($\Delta$) and standard deviation ($\sigma$) for all the pixels 
  are marked in each panel. Note that all the values have been enlarged by a factor of $10^4$.}
  \label{appfig:RF-model1}
\end{figure*}

\begin{figure*}
  \centering
  \includegraphics[width=16.8cm]{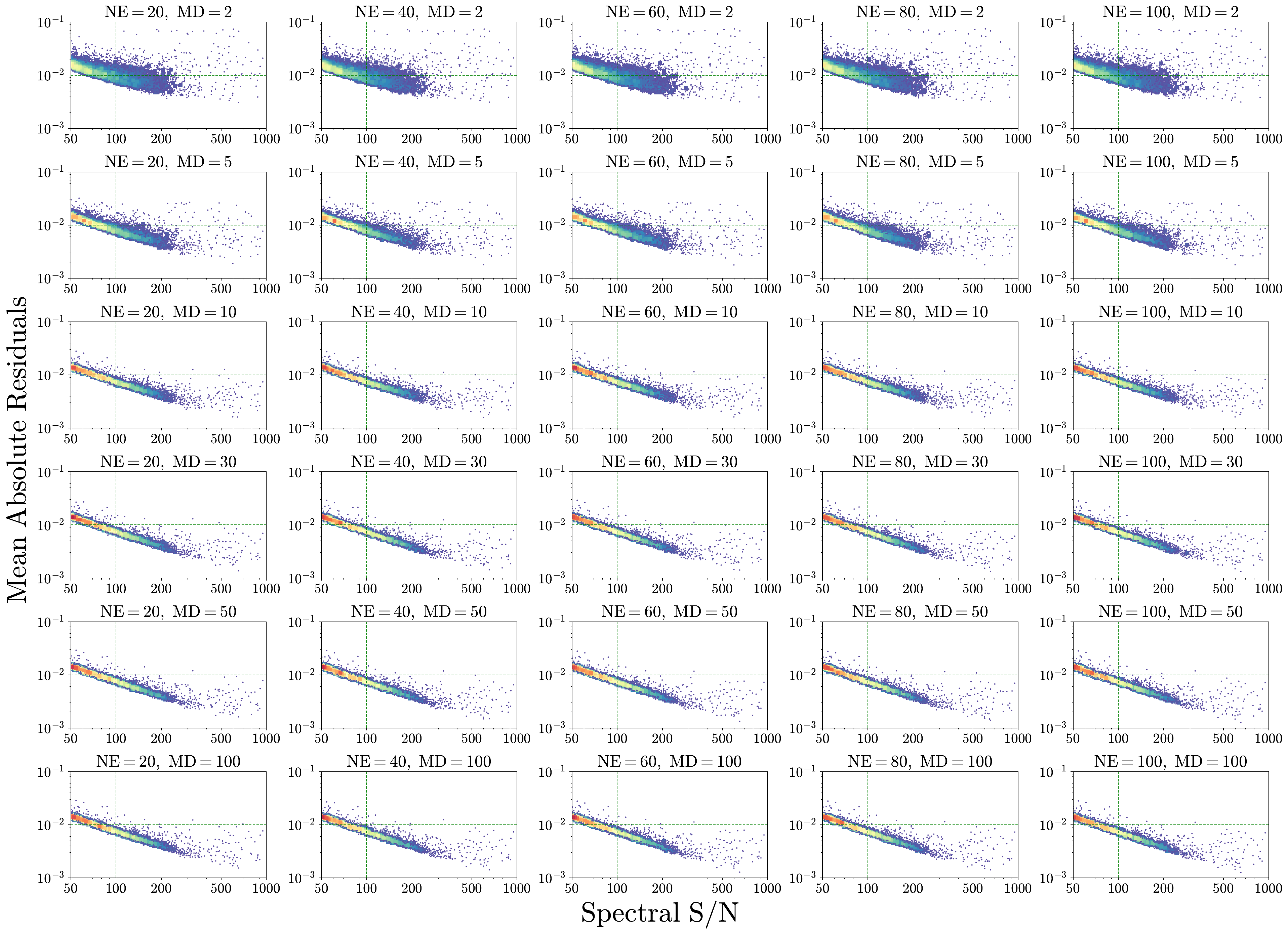}
  \caption{The mean of the absolute residuals (MAR) between observed and modeled normalized fluxes, taken along the wavelength in 
  the DIB window, for each RVS spectrum in the validation set (7324) as a function of the spectral S/N. The color represents 
  the number density of the spectra. The dashed green lines indicate $\rm MAR\,{=}\,0.01$ and $\rm S/N\,{=}\,100$ in each panel.}
  \label{appfig:RF-model2}
\end{figure*}

Figures \ref{appfig:RF-model1} and \ref{appfig:RF-model2} show the distribution of residuals between observed and modeled normalized 
fluxes of 7324 RVS spectra in the validation set. The performance of the validation set and the selection of the best RF model are 
discussed in detail in Sect. \ref{subsect:RF}.

\clearpage

\section{Injection test} \label{appsect:intest}

To validate our RF model and the DIB fittings, we did an injection test following the principles in \citetalias{Saydjari2023}. For
each observed RVS spectrum with $\rm S/N\,{>}\,50$ in the testing set (a total of 7324), we added a pair of synthetic DIB signals 
for both $\lambda$8621 and $\lambda$8648. The DIB parameters were uniformly sampled: $\mathcal{D}_{8621}\,{\in}\,[0.01,\,0.15]$, 
$\lambda_{8621}\,{\in}\,[8618,\,8628]$\,{\AA}, $\sigma_{8621} \in [0.6,\,3.0]$\,{\AA}, $\lambda_{8648} \in [8640,\,8650]$\,{\AA}, 
$\sigma_{8648} \in [8,\,12]$\,{\AA}, except for $\mathcal{D}_{8648}$ which was fixed as halved $\mathcal{D}_{8621}$. Then the ISM
spectra were derived from the observed spectra (plus synthetic DIBs) and the stellar templates predicted by the RF model. Finally,
we fitted the two DIBs with the same model and methods for real target spectra (Sect. \ref{subsect:dib-fit}). We also computed
Z-scores, defined as (fitted--injected values)/reported uncertainty, to quantify the bias of the fitted values and the uncertainties.

\begin{figure*}
  \centering
  \includegraphics[width=16.8cm]{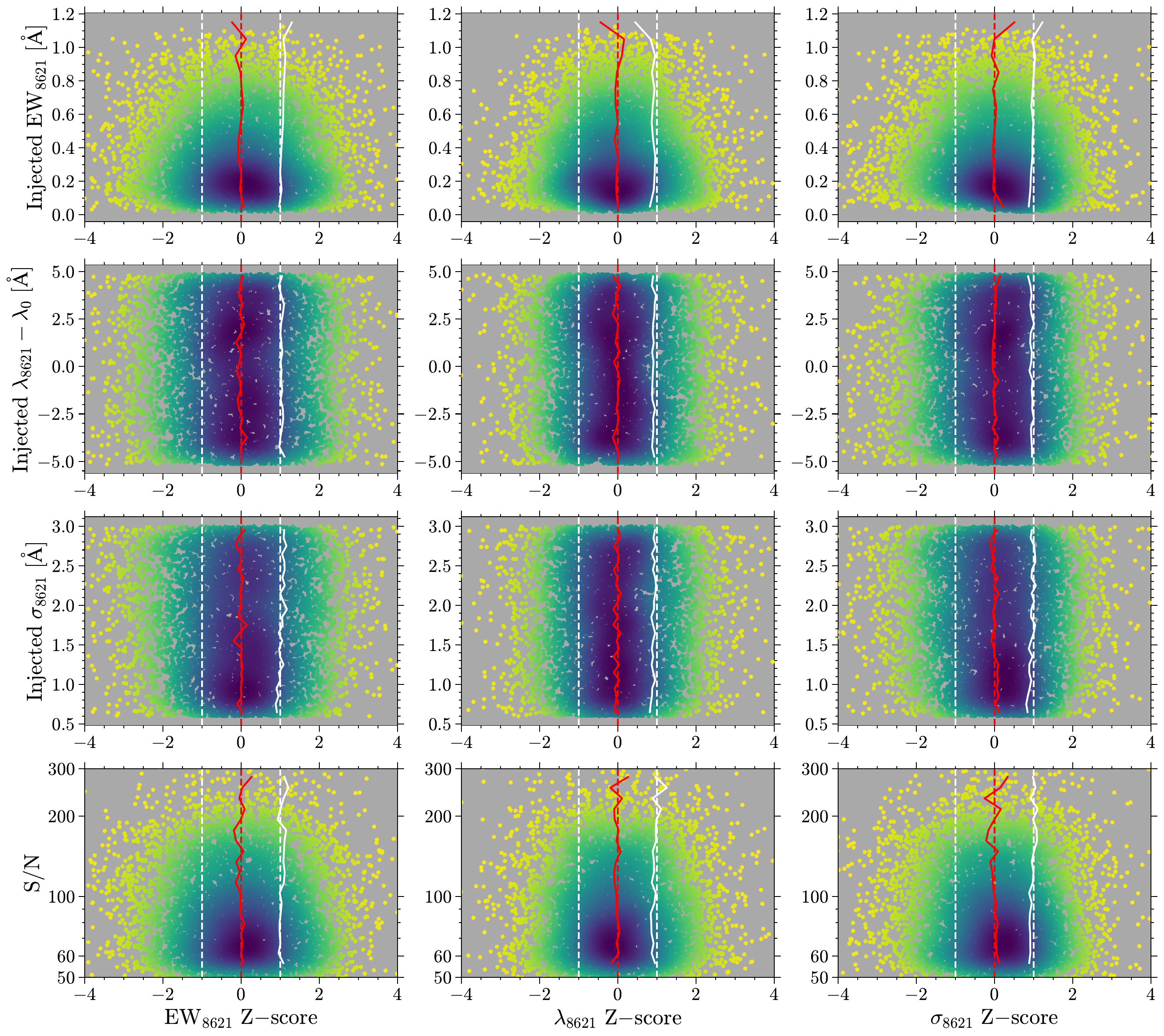}
  \caption{Distributions of the Z-scores computed from the injection tests of $\rm EW_{8621}$, $\lambda_{8621}$, and $\sigma_{8621}$,
  as a function of the injected DIB parameters and the S/N of the ISM spectra. The color scale indicates the number density estimated
  by the Gaussian KDE. In each panel, the solid red line indicates the variation of the mean differences of Z-scores with the injected
  DIB parameters and the S/N. The solid white line indicates the mean absolute differences of Z-scores. Dashed red and white
  lines provide a reference for Z-score equals 0 and $\pm$1.}
  \label{appfig:intest8621}
\end{figure*}

The distribution of $\rm EW_{8621}$ Z-score is almost perfectly uniform with respect to the injected parameters ($\rm EW_{8621}$,
$\lambda_{8621}$, and $\sigma_{8621}$) and the S/N of ISM spectra (see Fig. \ref{appfig:intest8621}), with a slight bias at the end
of injected $\rm EW_{8621}$ and S/N. Such distributions are also found for the Z-score of $\lambda_{8621}$ and $\sigma_{8621}$, and
they are highly consistent with the results in \citetalias{Saydjari2023}. The larger shakes of the mean differences varying with
injected DIB parameters and S/N (solid red lines in Fig. \ref{appfig:intest8621}) compared to \citetalias{Saydjari2023} should be
due to the smaller sample used in this work for the injection test. Because of the higher S/N cut ($\rm S/N\,{\geqslant}\,50$) for
the testing set, we do not capture the bias of Z-score at low S/N reported in \citetalias{Saydjari2023}. Further, the positive
bias of $\sigma_{8621}$ at large injected $\rm EW_{8621}$ is not as strong as in \citetalias{Saydjari2023}, but instead, there are
tiny negative biases of $\lambda_{8621}$ and $\sigma_{8621}$ with respect to $\rm EW_{8621}$ and S/N.

With the increase of injected $\rm EW_{8621}$, the mean absolute difference (MAD) of $\rm EW_{8621}$ Z-score (solid white line in 
Fig. \ref{appfig:intest8621}) becomes larger than 1, indicating an underestimation of the uncertainty of $\rm EW_{8621}$. The rising 
$\Delta \rm EW_{8621}$ is also found in the comparison between fitted and integrated $\rm EW_{8621}$, which is caused by the increasing 
noise and stellar residuals in low-S/N ISM spectra (see Sect. \ref{subsect:catalog} and Fig. \ref{fig:fit-int}). In the injection 
test, DIB features were randomly added to the observed spectra, while in fact, strong DIB signals are usually found in the spectra 
with generally lower S/N. Thus, the underestimation of the $\rm EW_{8621}$ uncertainty would be heavier for the target spectra. In 
contrast, the uncertainties of $\lambda_{8621}$ and $\sigma_{8621}$ seem to be overestimated, considering their Z-score variations 
with injected parameters.

\begin{figure*}
  \centering
  \includegraphics[width=16.8cm]{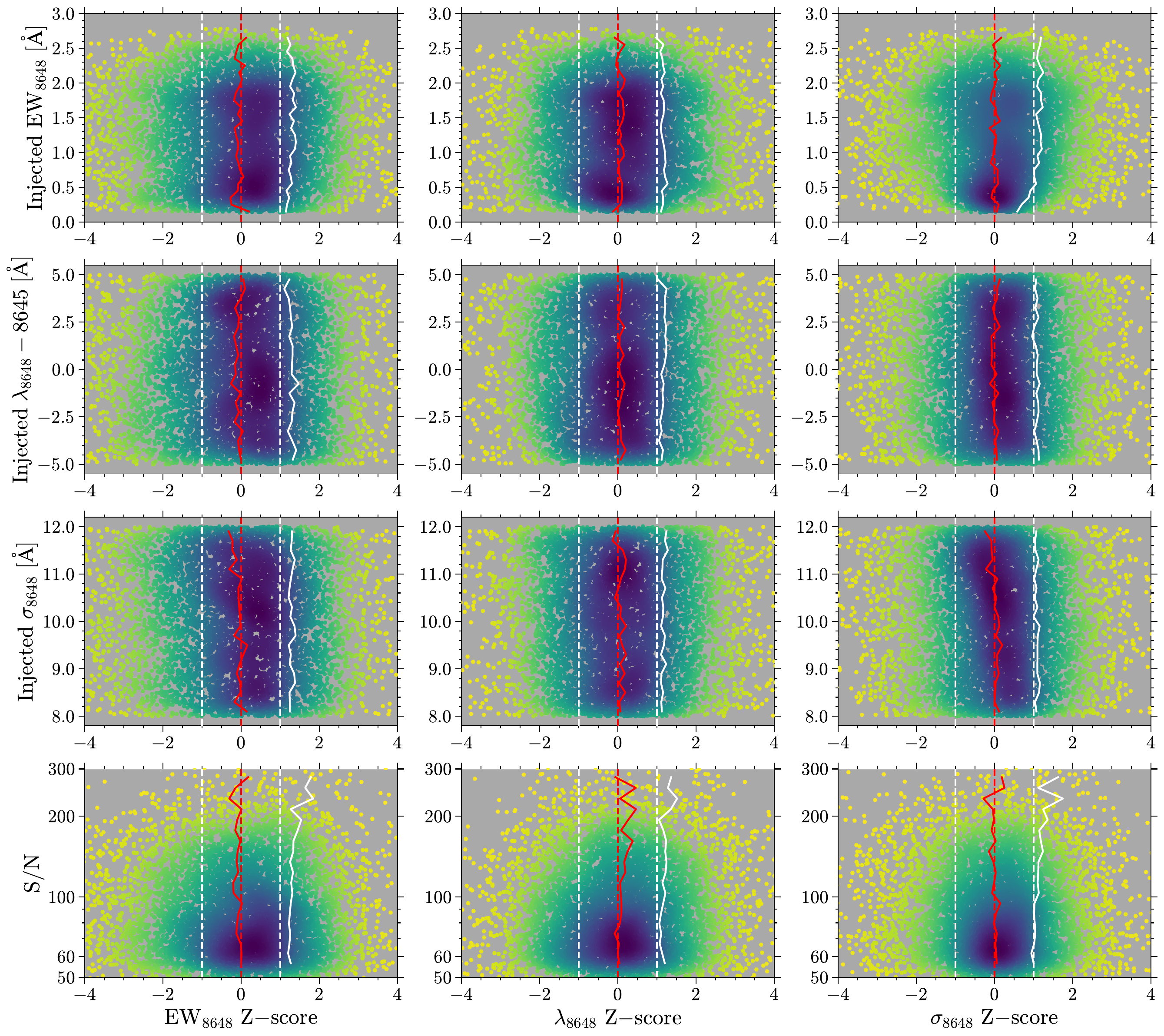}
  \caption{The same as Fig. \ref{appfig:intest8621}, but for DIB\,$\lambda$8648. }
  \label{appfig:intest8648}
\end{figure*}

The injection test was also applied to $\lambda$8648 (Fig. \ref{appfig:intest8648}). Compared to $\lambda$8621, stronger negative 
biases are found for $\rm EW_{8648}$, as well as positive biases for $\lambda_{8648}$. Moreover, the uncertainties of $\rm EW_{8648}$, 
$\lambda_{8648}$, and $\sigma_{8648}$ are clearly underestimated, and the magnitude of the overestimation tends to increase with 
S/N. The reason would be that the fitting of $\lambda$8648 is more easily affected by the correlated noise in its very broad profile.

Overall, we got similar distributions of the Z-scores with respect to the injected DIB parameters and S/N to those in 
\citetalias{Saydjari2023}, primarily verifying reliability and accuracy in our DIB fitting. The biases in the fitting of 
DIB\,$\lambda$8621 are tiny for $\rm EW_{8621}$, $\lambda_{8621}$, and $\sigma_{8621}$. The reported errors drawn from the MCMC
samplings can successfully describe the uncertainties of $\lambda_{8621}$ and $\sigma_{8621}$ (a bit overestimation), while the
reported uncertainty for $\rm EW_{8621}$ would be slightly underestimated (5\%--10\%). The fitting to $\lambda$8648 contains 
stronger biases and larger underestimated uncertainties due to the fact that this shallow and broad DIB is more difficult to be
fitted than $\lambda$8621. The injection test is still simple and ideal as we made use of a set of spectra with high S/N ($\geqslant$50) 
and fixed the ratio of $\mathcal{D}_{8621}/\mathcal{D}_{8648}\,{=}\,2$.

\clearpage

\section{The 77 DIB measurements for \texorpdfstring{$\,\lambda_0$}{l0} determination} \label{appsect:lambda0}

\begin{table*}[htpb]
\centering
\caption{The information of the 77 selected DIB measurements for the determination of $\lambda_0$ for DIB\,$\lambda8621$,
see Sect. \ref{subsect:kinematics} for details. }
\label{apptab:lambda0}
\begin{tabular}{crrrrcccc}
\hline\hline 
Line No. & Gaia-DR3 source ID & $\ell$ & $b$ & $G$ (mag) & $\teff$ (K) & $\logg$ & $\meta$ (dex) & $\lambda_{\rm helio}$ ({\AA}) \\ [0.5ex]
\hline   
1  & 3449303770318455552 & 175.08 &   1.11 & 10.61 & 4080.0 & 0.98 & --0.44 & $8623.34\,{\pm}\,0.20$ \\      
2  & 3450087202413909888 & 179.41 &   3.70 & 11.70 & 4568.0 & 1.92 & --0.09 & $8623.37\,{\pm}\,0.30$ \\      
3  & 3449953856568000128 & 179.89 &   4.65 & 11.35 & 4262.0 & 1.07 & --0.53 & $8623.10\,{\pm}\,0.15$ \\      
4  & 3449968631255768704 & 179.88 &   4.04 & 10.74 & 4893.0 & 2.44 & --0.10 & $8624.02\,{\pm}\,0.20$ \\      
5  & 3450047345121279872 & 179.51 &   3.20 & 11.41 & 4250.0 & 1.50 & --0.56 & $8623.74\,{\pm}\,0.15$ \\      
6  & 3404604666980354688 & 183.36 & --4.49 & 10.40 & 4537.0 & 1.55 & --1.30 & $8623.78\,{\pm}\,0.30$ \\      
7  & 3404613359994156160 & 183.30 & --4.33 & 12.35 & 4966.0 & 2.17 & --0.17 & $8623.76\,{\pm}\,0.25$ \\      
8  & 3404648273785638784 & 183.23 & --3.87 & 10.40 & 4515.0 & 1.66 & --0.29 & $8623.61\,{\pm}\,0.20$ \\      
9  & 3404713041892312448 & 182.93 & --3.95 & 11.23 & 4369.0 & 1.88 & --0.16 & $8622.95\,{\pm}\,0.35$ \\      
10 & 3429108181256596224 & 183.55 & --0.12 & 12.62 & 4250.0 & 1.50 & --0.04 & $8623.32\,{\pm}\,0.20$ \\      
11 & 3429156869005794816 & 183.13 & --0.90 & 11.14 & 4005.0 & 1.14 & --0.18 & $8622.95\,{\pm}\,0.20$ \\      
12 & 3429180607287003904 & 183.03 & --0.38 & 13.16 & 4506.0 & 1.44 & --0.28 & $8622.79\,{\pm}\,0.40$ \\      
13 & 3429205659834140160 & 182.80 & --0.24 & 12.88 & 3770.0 & 0.49 & --0.49 & $8623.87\,{\pm}\,0.35$ \\      
14 & 3427983582721810816 & 185.30 &   0.19 & 11.11 & 3949.0 & 0.90 & --0.47 & $8622.76\,{\pm}\,0.20$ \\      
15 & 3429919242882519296 & 183.99 &   1.22 & 11.00 & 3833.0 & 0.59 & --0.46 & $8622.80\,{\pm}\,0.20$ \\      
16 & 3429812482875636864 & 184.18 &   0.77 & 11.56 & 4057.0 & 0.89 & --0.76 & $8622.81\,{\pm}\,0.25$ \\      
17 & 3428490461880542720 & 183.61 & --3.04 & 12.98 &    --- &  --- &    --- & $8623.77\,{\pm}\,0.30$ \\      
18 & 3429840902678074496 & 183.88 &   1.07 & 11.34 & 3836.0 & 0.40 & --0.80 & $8623.83\,{\pm}\,0.25$ \\      
19 & 3428112225582653952 & 184.54 & --1.18 & 11.55 & 3854.0 & 0.58 & --0.51 & $8623.70\,{\pm}\,0.20$ \\      
20 & 3428170534058648704 & 184.23 & --1.15 & 12.22 & 5103.0 & 2.85 & --0.48 & $8623.36\,{\pm}\,0.30$ \\      
21 & 3428189126975132288 & 183.95 & --1.22 & 11.04 & 3943.0 & 0.89 & --0.62 & $8623.30\,{\pm}\,0.15$ \\      
22 &  181938353713330432 & 172.12 & --2.98 & 11.76 & 4996.0 & 2.60 & --0.17 & $8623.81\,{\pm}\,0.30$ \\      
23 &  181316201928996992 & 173.40 & --4.34 & 12.15 & 4074.0 & 1.03 & --0.34 & $8623.39\,{\pm}\,0.25$ \\      
24 &  180928417922548352 & 174.12 & --2.11 & 12.09 & 4250.0 & 1.50 & --2.01 & $8622.52\,{\pm}\,0.20$ \\      
25 &  181012324403023232 & 173.62 & --3.26 & 12.70 & 4250.0 & 1.50 & --2.00 & $8623.34\,{\pm}\,0.25$ \\      
26 &  183692388293795968 & 171.97 & --0.37 & 12.28 & 4337.0 & 1.17 & --0.23 & $8623.20\,{\pm}\,0.25$ \\      
27 &  183707674080750464 & 172.10 &   0.02 & 11.34 & 3254.0 & 4.40 & --0.73 & $8623.23\,{\pm}\,0.15$ \\      
28 &  184504201534107520 & 170.03 &   0.84 & 11.53 & 3774.0 & 0.47 & --0.83 & $8623.17\,{\pm}\,0.20$ \\      
29 &  182710004719137536 & 173.06 & --1.25 & 11.30 & 4056.0 & 0.94 & --0.45 & $8623.29\,{\pm}\,0.25$ \\      
30 & 3377490370941037568 & 188.70 &   3.50 & 12.88 & 4250.0 & 1.50 & --0.32 & $8624.32\,{\pm}\,0.35$ \\      
31 & 3421535909099285760 & 177.85 & --4.83 & 10.09 &    --- &  --- &    --- & $8623.09\,{\pm}\,0.25$ \\      
32 & 3423369142877208960 & 188.90 & --0.32 & 10.30 &    --- &  --- &    --- & $8623.52\,{\pm}\,0.45$ \\      
33 & 3427376656600676736 & 185.27 & --2.41 & 11.09 & 4250.0 & 1.50 & --0.65 & $8624.09\,{\pm}\,0.15$ \\      
34 & 3426712551578699008 & 186.39 &   3.33 & 10.89 & 4007.0 & 1.01 & --0.29 & $8623.45\,{\pm}\,0.25$ \\      
35 & 3426946949415969664 & 185.34 &   4.03 & 10.35 & 7425.0 & 3.89 & --3.85 & $8623.47\,{\pm}\,0.35$ \\      
36 & 3374812162475721472 & 189.85 & --0.22 & 10.89 & 3950.0 & 0.86 & --0.30 & $8623.58\,{\pm}\,0.20$ \\      
37 & 3444339269159283712 & 179.21 &   0.17 &  9.99 & 4749.0 & 2.39 & --0.13 & $8624.20\,{\pm}\,0.25$ \\      
38 & 3443660556949845760 & 180.35 &   1.44 &  9.36 &    --- &  --- &    --- & $8623.77\,{\pm}\,0.25$ \\      
39 & 3443304422559206656 & 180.55 & --0.36 & 10.23 &    --- &  --- &    --- & $8622.79\,{\pm}\,0.25$ \\      
40 & 3443076269598866432 & 181.21 & --0.17 & 10.84 & 4774.0 & 2.03 & --0.29 & $8623.21\,{\pm}\,0.40$ \\ [0.5ex]
\hline
\end{tabular}
\tablefoot{Columns: Line number, Gaia-DR3 source ID, Galactic longitude, Galactic latitude, apparent $G$-band
magnitude, stellar atmospheric parameters ($\teff$, $\logg$, $\meta$) from {\gspspec}, and the central wavelength 
of DIB\,$\lambda$8621 in the heliocentric frame ($\lambda_{\rm helio}$) with the mean error from the lower and 
upper confidence levels of $\lambda_{8621}$.}
\end{table*}\addtocounter{table}{-1}

\begin{table*}[htpb]
\centering
\caption{--continued.}
\begin{tabular}{crrrrcccc}
\hline\hline 
Line No. & Gaia-DR3 source ID & $\ell$ & $b$ & $G$ (mag) & $\teff$ (K) & $\logg$ & $\meta$ (dex) & $\lambda_{\rm helio}$ ({\AA}) \\ [0.5ex]
\hline   
41 & 3443784565544612864 & 180.24 &   2.40 &  9.13 &    --- &  --- &    --- & $8623.53\,{\pm}\,0.25$ \\
42 & 3444677987460354688 & 177.83 &   0.42 & 10.21 &    --- &  --- &    --- & $8623.99\,{\pm}\,0.20$ \\
43 & 3444197878835192832 & 179.30 &   0.17 & 11.07 & 4580.0 & 1.67 & --0.42 & $8623.34\,{\pm}\,0.25$ \\
44 & 3444204991301278336 & 179.34 & --0.59 & 10.39 &    --- &  --- &    --- & $8623.43\,{\pm}\,0.20$ \\
45 & 3444895278446243456 & 178.67 &   2.74 &  9.93 &    --- &  --- &    --- & $8623.62\,{\pm}\,0.45$ \\
46 &  189823952326071296 & 171.89 &   4.44 & 10.19 &    --- &  --- &    --- & $8624.04\,{\pm}\,0.35$ \\
47 & 3441341794303524352 & 181.93 & --1.71 &  9.02 &    --- &  --- &    --- & $8623.55\,{\pm}\,0.20$ \\
48 & 3441819978782431488 & 180.43 & --1.21 & 11.69 & 4770.0 & 2.12 & --0.26 & $8623.35\,{\pm}\,0.30$ \\
49 & 3441427723712935936 & 181.59 & --1.45 &  9.36 &    --- &  --- &    --- & $8623.72\,{\pm}\,0.25$ \\
50 & 3442367359479512832 & 179.19 & --2.65 & 12.43 & 4250.0 & 1.50 & --0.77 & $8623.07\,{\pm}\,0.15$ \\
51 & 3441921099492825216 & 180.68 & --3.51 &  9.38 &    --- &  --- &    --- & $8623.55\,{\pm}\,0.20$ \\
52 & 3398771349776851712 & 189.19 & --2.81 & 10.99 & 4046.0 & 0.89 & --0.56 & $8623.96\,{\pm}\,0.40$ \\
53 & 3399714146637272192 & 188.01 & --3.81 & 12.13 & 4006.0 & 0.83 & --0.55 & $8623.90\,{\pm}\,0.20$ \\
54 & 3399768812980401408 & 187.69 & --3.79 & 11.94 & 4277.0 & 1.44 & --0.32 & $8623.22\,{\pm}\,0.25$ \\
55 & 3399775405754199296 & 187.67 & --3.68 & 10.41 & 7840.0 & 4.31 &    --- & $8624.15\,{\pm}\,0.40$ \\
56 & 3431540472779533312 & 181.43 &   1.74 &  9.92 &    --- &  --- &    --- & $8623.87\,{\pm}\,0.30$ \\
57 & 3430697250437513600 & 182.97 &   0.35 & 12.87 & 4250.0 & 1.50 & --0.39 & $8623.37\,{\pm}\,0.25$ \\
58 & 3430721469754522624 & 182.62 &   0.27 & 11.21 & 4107.0 & 1.23 & --0.38 & $8624.02\,{\pm}\,0.20$ \\
59 & 3447715761994147328 & 177.23 &   0.70 & 12.08 & 4109.0 & 1.04 & --0.33 & $8622.82\,{\pm}\,0.30$ \\
60 & 3448966696986298112 & 174.98 & --0.81 & 11.73 & 3894.0 & 0.58 & --0.42 & $8623.22\,{\pm}\,0.35$ \\
61 & 3448977352805311488 & 174.93 & --0.49 & 10.79 & 4027.0 & 0.14 & --0.85 & $8623.01\,{\pm}\,0.25$ \\
62 & 3448621145396695040 & 175.39 &   2.10 &  9.58 &    --- &  --- &    --- & $8623.80\,{\pm}\,0.25$ \\
63 & 3447868564045571328 & 176.66 & --0.78 & 10.39 & 4246.0 & 1.29 & --0.42 & $8623.61\,{\pm}\,0.35$ \\
64 & 3448730301988310016 & 175.46 & --0.99 & 11.64 & 4019.0 & 1.08 & --0.27 & $8623.63\,{\pm}\,0.35$ \\
65 & 3447921336308217728 & 176.30 & --0.36 & 12.18 & 4005.0 & 0.70 & --0.37 & $8623.07\,{\pm}\,0.25$ \\
66 & 3447966794242406016 & 176.73 &   0.42 & 11.92 & 3800.0 & 0.12 & --0.84 & $8623.71\,{\pm}\,0.20$ \\
67 & 3448018058969624960 & 176.25 & --0.13 & 11.85 & 3728.0 & 0.75 & --0.58 & $8624.02\,{\pm}\,0.20$ \\
68 & 3448028817865787904 & 176.07 &   0.11 &  8.80 &    --- &  --- &    --- & $8623.47\,{\pm}\,0.30$ \\
69 & 3448041629750130560 & 176.15 &   0.34 & 11.67 & 3721.0 & 0.48 & --0.54 & $8623.36\,{\pm}\,0.25$ \\
70 & 3449201412658292352 & 175.59 &   0.71 & 11.21 &    --- &  --- &    --- & $8623.36\,{\pm}\,0.20$ \\
71 & 3454808750160408320 & 174.88 &   3.76 &  9.76 &    --- &  --- &    --- & $8623.44\,{\pm}\,0.20$ \\
72 & 3455866652144528000 & 173.41 &   3.52 &  9.29 &    --- &  --- &    --- & $8623.32\,{\pm}\,0.25$ \\
73 & 3454900009626360192 & 175.73 &   4.21 & 11.85 & 3970.0 & 0.48 & --0.69 & $8622.35\,{\pm}\,0.30$ \\
74 & 3455911079286056320 & 173.00 &   3.27 & 10.97 & 4008.0 & 0.90 & --0.40 & $8623.15\,{\pm}\,0.20$ \\
75 & 3455913587546936448 & 172.94 &   3.34 &  9.54 & 4542.0 & 2.01 & - 0.12 & $8623.26\,{\pm}\,0.20$ \\
76 & 3454549089320508032 & 175.95 &   2.54 &  9.38 &    --- &  --- &    --- & $8623.42\,{\pm}\,0.15$ \\
77 & 3456281026291047936 & 173.16 &   4.59 &  9.83 &    --- &  --- &    --- & $8622.80\,{\pm}\,0.30$ \\ [0.5ex]
\hline
\end{tabular}
\end{table*}

\clearpage

\section{Selection effect on the DIB catalog} \label{appsect:select-bias}

\begin{figure*}
\centering
\includegraphics[width=16.8cm]{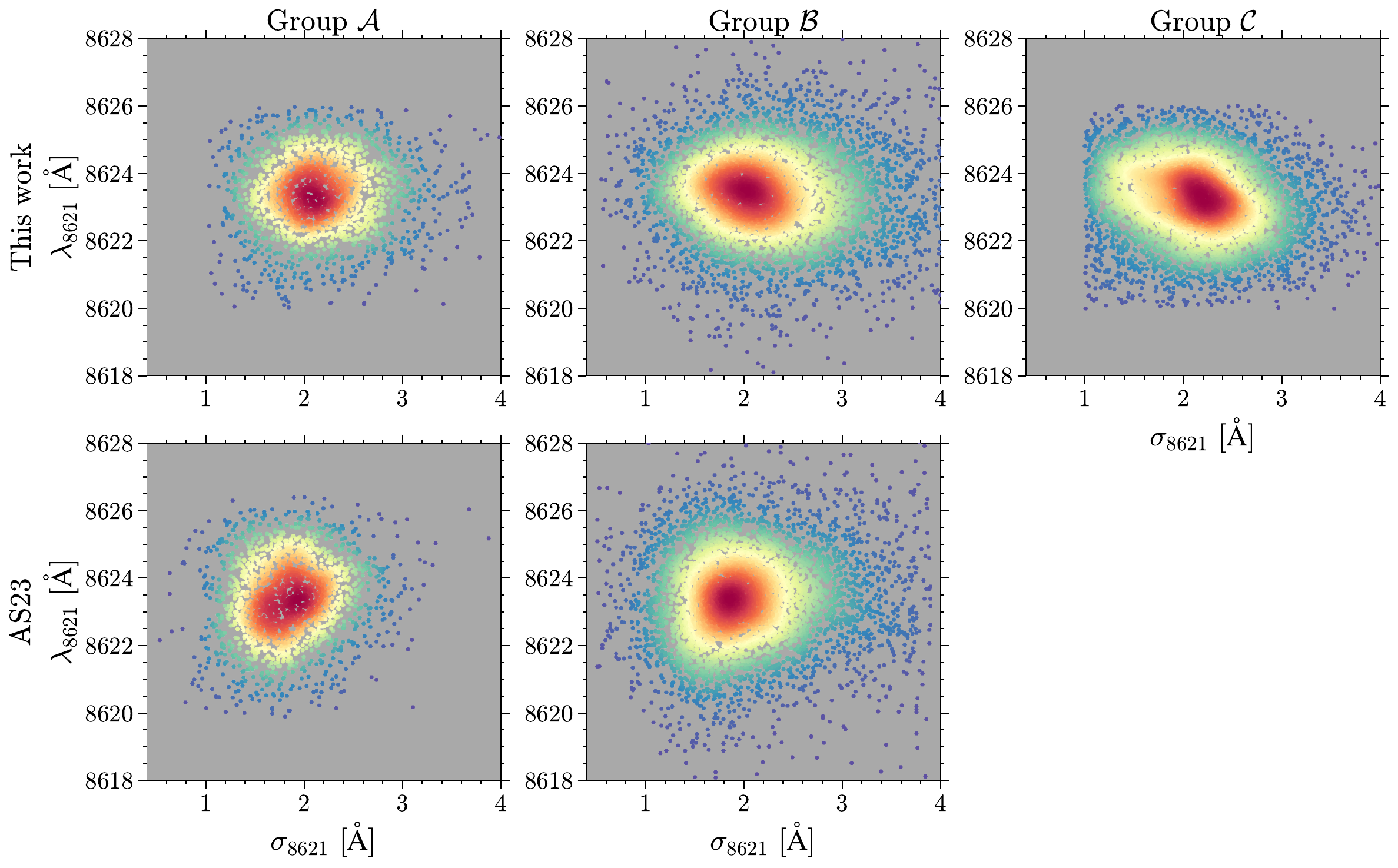}
\caption{The $\sigma_{8621}-\lambda_{8621}$ distributions for different groups defined in Appendix \ref{appsect:select-bias}.
The DIB parameters in the first row come from this work, and those in the second row come from \citetalias{Saydjari2023}.
The color of the scattered points represents their number density estimated by the Gaussian KDE.} 
\label{fig:complete1} 
\end{figure*}

\begin{figure*}
\centering
\includegraphics[width=16.8cm]{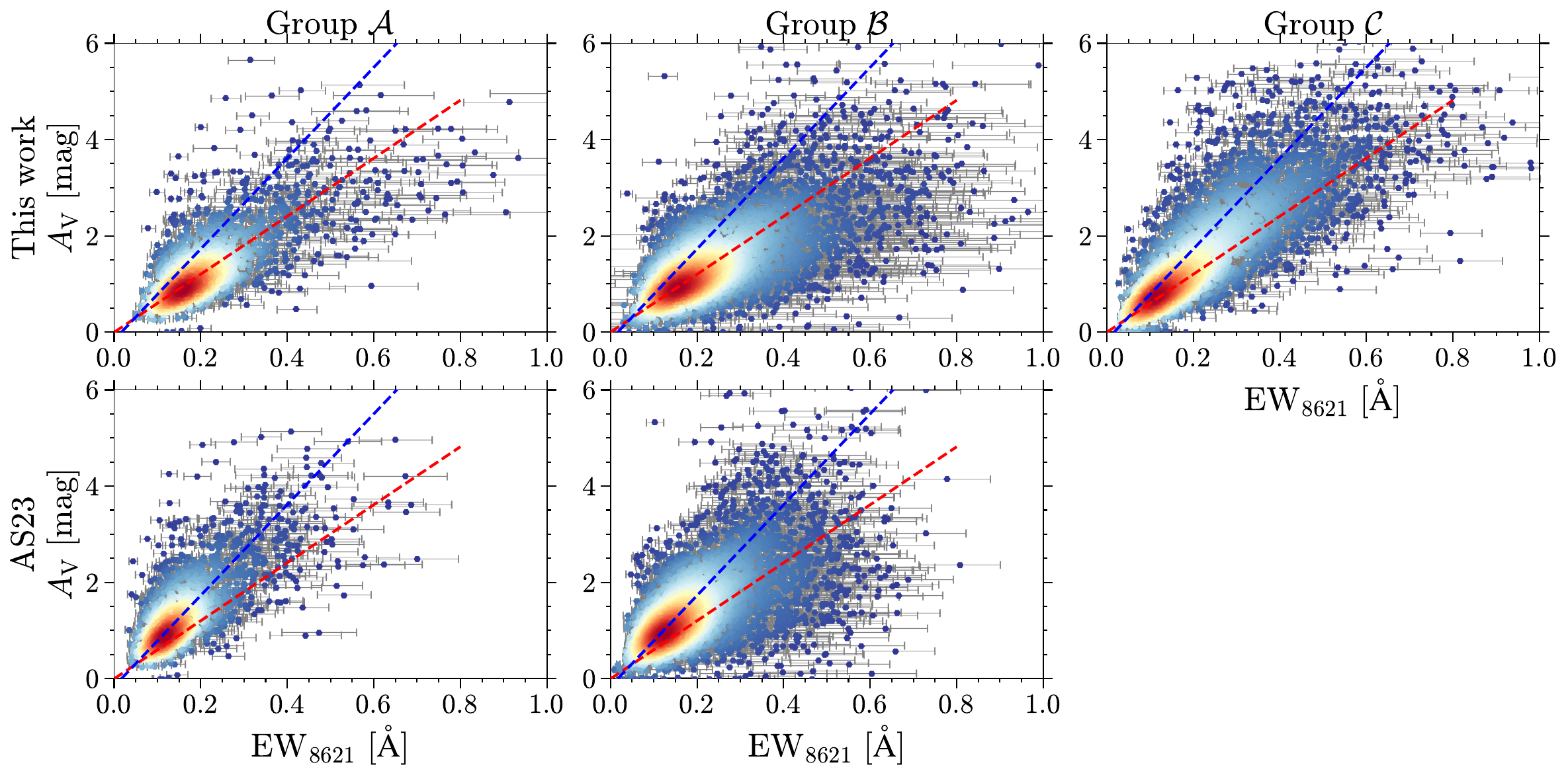}
\caption{Correlation between $\rm EW_{8621}$ and $\Av$ for different groups defined in Appendix \ref{appsect:select-bias}.
The DIB parameters in the first row come from this work, and those in the second row come from \citetalias{Saydjari2023}.
The color of the scattered points represents their number density estimated by the Gaussian KDE. The dashed red and blue 
lines show the linear fits to $\rm EW_{8621}$ and $\Av$ from \citetalias{Schultheis2023DR3} and \citetalias{Saydjari2023}, 
respectively.} 
\label{fig:complete2} 
\end{figure*}

Besides the systematic differences in DIB parameters (see Sect. \ref{subsect:GDR3}), different selection criteria between this work 
and \citetalias{Saydjari2023} also result in very different selected DIB catalogs, that is their joint group (2000 cases) only
accounts for $\sim$25\% of the total for each. There cannot be so many DIB signals detecting in one work but disappearing in another.
Except $\chi^2$ used in both studies, our cuts are based on the $\mathcal{D}/R_C$ and S/N of the ISM spectra to control the noise
level, as well as additional constraints on $\lambda_{8621}$ and $\sigma_{8621}$. While the cuts in \citetalias{Saydjari2023} are 
based on the DIB SNR defined by $\Delta \chi^2$ and the eigenvector in their model. We investigate the selection effect on the two
DIB catalogs by comparing different crossed groups. Specifically, Group $\mathcal{A}$ contains 2000 DIB measurements detected 
in both \citetalias{Saydjari2023} and this work, Group $\mathcal{B}$ contains 5463 DIB measurements detected in \citetalias{
Saydjari2023} but not in this work, and Group $\mathcal{C}$ contains 5619 DIB measurements detected not in \citetalias{Saydjari2023} 
but in this work. Here we only consider 7463 DIB measurements in \citetalias{Saydjari2023} whose background stars are 
contained in our target sample. The $\sigma_{8621}-\lambda_{8621}$ distribution and the ${\rm EW_{8621}}-\Av$ correlation for 
different groups are shown in Figs. \ref{fig:complete1} and \ref{fig:complete2}, respectively.

The $\sigma_{8621}$ and $\lambda_{8621}$ in this work presents a compact Gaussian distribution for Group $\mathcal{A}$, but a clear
contamination for small $\sigma_{8621}$ can be seen for Group $\mathcal{C}$. Although we cannot access the dropped cases in
\citetalias{Saydjari2023} (no data for Group $\mathcal{C}$), for their full sample shown in Fig. 21 in \citetalias{Saydjari2023},
there seems to be a contamination for $\sigma_{8621}\,{\lesssim}\,1$\,{\AA} as well, but those cases lead to a smaller $\lambda_{8621}$
in contrast to this work. Group $\mathcal{B}$, with dropped measurements in this work, also presents a good $\sigma_{8621}-\lambda_{8621}$ 
distribution, which clearly shows the selection bias of our DIB catalog. For \citetalias{Saydjari2023}, Group $\mathcal{B}$
shows a better $\sigma_{8621}-\lambda_{8621}$ distribution than Group $\mathcal{A}$ which contains a weak distortion, while Group 
$\mathcal{B}$ contains measurements with large $\sigma_{8621}\,{\gtrsim}\,3$\,{\AA}. The weak distortion could be the reason for
the variation of $\Delta \lambda_{8621}$ seen in Fig. \ref{fig:GDR3} between \citetalias{Saydjari2023} and this work. Comparing 
Groups $\mathcal{B}$ and $\mathcal{C}$, the cuts applied in \citetalias{Saydjari2023} can exclude the noisy cases with small 
$\sigma_{8621}$, caused by the observational noise and/or the stellar residuals, more efficiently than ours. 

For the ${\rm EW_{8621}}-\Av$ correlation, the high-density regions (from red to yellow in Fig. \ref{fig:complete2}) present a
consistent tendency among different groups for this work and \citetalias{Saydjari2023}, respectively, with a similar magnitude of
dispersion, despite of a systematic difference in mean ${\rm EW_{8621}}/\Av$ ratio between \citetalias{Saydjari2023} and this work.
The fewer scatters in Groups $\mathcal{A}$ and $\mathcal{C}$ demonstrate that our cuts can exclude more extreme cases with small
$\Av$ but large $\rm EW_{8621}$ than \citetalias{Saydjari2023}.

All groups with some selection criteria can present proper properties for the DIB measurements (the $\sigma_{8621}-\lambda_{8621}$ 
distribution and the ${\rm EW_{8621}}-\Av$ correlation considered in this analysis) and they are self-consistent for one work with
its own measurements. Hence, present DIB catalogs built in different studies would have high purity but low completeness. A catalog 
with more cuts (e.g. the joint sample between \citetalias{Saydjari2023} and this work) will obtain more accurate DIB measurements 
but at the expense of sample size. 

\clearpage

\end{CJK*}

\end{document}